\documentclass[aps,amssymb,floatfix,prd,preprintnumbers]{revtex4-2}
\usepackage{epstopdf}
\usepackage{capt-of}
\usepackage{graphicx}  
\usepackage{dcolumn}   
\usepackage{bm}
\usepackage{amsmath}
\usepackage{color}
\usepackage[font=scriptsize]{caption}
\usepackage{subfigure}
\usepackage[colorlinks]{hyperref}
\begin{document}
\input epsf.tex
\title{\bf Matter Bounce Scenario and the Dynamical aspects in $f(Q,T)$ Gravity}

\author{A.S. Agrawal\footnote{Department of Mathematics, Birla Institute of Technology and Science-Pilani, Hyderabad Campus, Hyderabad-500078, India}, Laxmipriya Pati\footnote{Department of Mathematics, Birla Institute of Technology and Science-Pilani, Hyderabad Campus, Hyderabad-500078, India}, S.K. Tripathy\footnote{Department of Physics, Indira Gandhi Institute of Technology, Sarang, Dhenkanal, Odisha-759146, India, tripathy\_sunil@rediffmail.com}, B. Mishra\footnote{Department of Mathematics, Birla Institute of Technology and Science-Pilani, Hyderabad Campus, Hyderabad-500078, India, email: bivu@hyderabad.bits-pilani.ac.in}}

\affiliation{ }

\begin{abstract}
\textbf{Abstract}
In the context of the late time cosmic acceleration phenomenon, many geometrically modified theories of gravity have been proposed in recent times. In this paper, we have investigated the role of a recently proposed extension of symmetric teleparallel gravity dubbed as $f(Q,T)$ gravity in getting viable cosmological models, where $Q$  and $T$ respectively denote the non-metricity and the trace of energy momentum tensor.  We stress upon the mathematical simplification of the formalism in the $f(Q,T)$ gravity and derived the dynamical parameters in more general form in terms of the Hubble parameter.  We considered two different cosmological models mimicking non-singular matter bounce scenario. Since energy conditions play a vital role in providing bouncing scenario, we have analyzed different possible energy conditions to show that strong energy condition and null energy condition be violated in this theory. The models considered in the work are validated through certain cosmographic tests and stability analysis.

\end{abstract}

\maketitle

\section{Introduction}
It is well known that General Relativity (GR) has been  formulated in the Riemann metric space and basically it is a geometric theory. A satisfactory description of the gravitational field requires a more generalized theory of gravity involving more general geometric structures. So that at large scale, the proper explanation of the evolution of Universe can be revealed. This becomes necessary after a lot of cosmological observations provided strong evidences \cite{Riess98,Perlmutter99,Riess11,Ade16,Aghanim20} supporting the accelerated expansion phase of the Universe at a late phase of its evolution. The standard cosmological model or GR has a long history of successes in addressing many issues in cosmology and astrophysics but the limitations of GR in addressing issues like initial singularity, flatness, cosmological horizon, dark energy, dark matter necessitates the formulation of more generalised theories. The inflation scenario could address some of the issues partially, but the issue in early Universe, the initial singularity \cite{Borde96,Borde03} and in the present Universe the late time cosmic acceleration \cite{Riess98,Perlmutter99,Riess11,Ade16,Aghanim20} pose major challenges to GR as well to modern cosmology. It was Weyl  \cite{Weyl18} who created a more general geometry to make the geometrical unification of electromagnetism and gravitation. In this theory of Weyl, the covariant divergence of the metric tensor is non-zero. Mathematically this can be expressed in term of new geometric quantity, known as the non-metricity. This proposal of Weyl's could not get the attention during the initial days because of the criticism of Einstein that Weyl's theory is in contradiction with the known experimental outcomes \cite{Einstein18}. The second attempt was made by Cartan who proposed the Einstein-Cartan theory, which was an extension of GR \cite{Cartan23}. In this extension, the torsion field was proposed, which can be interpreted physically as the spin density. The Weyl geometry can be naturally extended to include the torsion. The third attempt made by Weitzenb$\ddot{o}$ck \cite{Weitzenbock23} known as Weitzenb$\ddot{o}$ck space, which was an independent mathematical development with important physical applications. The geometries have the property of teleparallelism as the Riemann curvature tensor of a Weitzenb$\ddot{o}$ck space remains zero. Einstein \cite{Einstein28} was first to apply this space-time and proposed the unified teleparallel theory of electromagnetism and gravity. The basic idea in the teleparallel gravity is to replace the metric $g_{\mu \nu}$ of the space-time by a set of tetrad vectors, $e^i_{\mu}$ and the torsion that generated from the tetrad fields provide the gravitational effects. Now, the curvature of the geometry replaced by the torsion, and hence known as the teleparallel equivalent of GR (TEGR)\cite{Moller61}. In fact, GR can be described in two geometrical representation such as (i) curvature representation, where the torsion and non-metricity are zero; and (ii) the teleparallel representation, where the curvature and non-metricity vanish. Another possible representation is the symmetric teleparallel gravity, where the geometric variable can be represented by the non-metricity $Q$ \cite{Nester99}. This has been further developed into the $f(Q)$ gravity \cite{Jiemenez18}. Again the $f(Q)$ gravity has been extended by framing the non-minimal coupling between non-metricity $Q$ and trace energy momentum tensor of matter $T$, to form $f(Q,T)$ gravity \cite{Xu19}. It is worthy to mention that the action of $f(T)$ and $f(Q)$ gravity respectively  $\int d^4x\sqrt{-g}T$ and $\int d^4x\sqrt{-g}Q$ are equivalent to GR in flat space \cite{Jimenez19}. However the generalization of $f(T)$ \cite{Wu10,Li11,Cai16,Benetti20} and $f(Q)$ gravity \cite{Lu19,Dialektopoulos19,Bajardi20,Jimenez20} can be ascribed as the modification in GR.\\

Of late, a lot of attention is being given to the $f(Q)$ and  $f(Q,T)$ gravity because of the agreement on the late time cosmic acceleration  and other issues of early Universe. To mention, the strong coupling problems occurred when the perturbation is considered in the FLRW background is absent in $f(Q)$ gravity \cite{Jimenez20} whereas it occurred in $f(T)$ gravity \cite{Golovnev18}. Also, the evolution of the matter density perturbation at the linear perturbation level has been tested with the redshift space distortion data \cite{Barros20}. The Lagrangian of $f(Q)$ can be reformulated as an explicit function of the redshift function $f(z)$ and with this redshift approach, several modified $f(Q)$ model observational analysis can be performed \cite{Lazkoz19}. On the non-metricity $Q$ based modified gravity theory, Frusciante has framed the theoretical predictions on linear cosmological observables \cite{Frusciante21}. Latorre et al. \cite{Latorre18} have shown the generic property theories with the non-metricity $Q$. In their seminal work Soudi et al. \cite{Soudi19} have explored the possibility of gravitational waves in the symmetric teleparallel gravity with their extension and have also shown the effect with a 
non-minimal coupling between $f(Q)$ and scalar field. The gravitation in this context has been expressed through the non-metricity in stead of the curvature or torsion. With observational constraints,  Zia et al. \cite{Zia21} have shown the transit cosmological model with an early deceleration and late time acceleration. Very recently, a mathematical formalism has been developed for the $f(Q,T)$ gravity in terms of the Hubble function  by Pati et al. \cite{Pati21}, so that the cosmological model concerning late time acceleration can be investigated with the known Hubble function. Najera and Fajardo \cite{Najera21} have shown the $\Lambda$CDM behaviour of the cosmological model with the constrained values of the parameters.  \\

Another issue that perturbs the cosmologists is the occurrence of initial singularity, which makes the physical laws presuppose the space-time. As a result, if singularity occurs, the description of the space-time breaks down \cite{Earman95,Barros98}. In the FLRW solution, the occurrence of initial singularity has become inevitable and Friedmann claimed its occurrence at the beginning of the evolution of Universe \cite{Friedmann22,Friedmann24}. Novello and Salim \cite {Novello79} and Melnikov and Orlov \cite{Melnikov79} were the first to provide the exact solutions  to the bouncing geometry. However, due to the belief that appearance of inflation might solve the singularity issue, their results did not get much attention.  In GR, to experience the non-singular bounce, the strong energy condition, $\rho+3p$ should be negative; the recent cosmological observations on accelerated expansion of the Universe \cite{Riess98,Perlmutter99} also  indicate the violation of strong energy condition i.e. $\rho+3p<0$ through the speculation of an exotic dark energy form within the purview of GR.  Therefore, the possibility of occurrence of non-singular bounce in the evolution of Universe again become the attraction of recent research. In addition, since the cosmological singularity is unavoidable in GR, it is necessary to go beyond GR or to introduce some new form of matter that violates some key energy conditions \cite{Hawking73}, such as the null energy condition. In this context, the singularity theorem of GR would not be applicable and the emergence of non-singular cosmological solution may be possible. Several geometrical proof are available in the literature to support this content \cite{Bohm87, Bojowald01,Khoury01,Erickson04, Feng06,Horava09,Lin11,Qiu11}. \\

Several discussions are available in literature based on the two emerging problems concerning the matter bounce scenario. One is the increase of energy density of radiation as the Universe contracts and the other one is the contracting pre bounce phase due to the presence of anisotropies  \cite{Guth81}. So, in the context of GR, the cosmological bounce can not be described in a homogeneous and isotropic background. In addition, when the kinetic energy dominates then the Universe may collapse to singularity and this can be avoided if the Universe expands before attaining singularity during contraction. This is possible in case of the bouncing Universe and possible if the bounce occurs before the energy comes up to the Planck scale, then GR can be used as an effective field theory \cite{Matsui19}. To know more on this discussions, one may refer \cite{Novello08,Battefeld15}. In connection with the cosmological observation on the relation between the bouncing behaviour and anomalies in CMB, some one can refer the work \cite{Liu13}. Post Supernovae findings, several attempts are being taken to realize the bouncing scenario in the extended theories such as, in $f(R)$ Gravity \cite{Cai11,Paul14,Bamba14,Bhattacharya16,Saridakis18,Ilyas21}, $f(R,T)$ gravity \cite{Shabani18,Mishra19,Tripathy21}, $f(R,G)$ gravity \cite{Elizalde20}, $f(T)$ gravity \cite{Logbo19,Skugoreva20}, $f(T,B)$ gravity \cite{Caruana20}. In fact, bouncing cosmology can be studied by modifying both the gravitational actions and the matter field \cite{Boisseau15}. Since these extended theories of gravity have been successful in resolving the bouncing scenario, we are motivated here to study the bouncing scenario in another geometrically modified gravity, the $f(Q,T)$ gravity, where $Q$ is the non-metricity and $T$ is the trace of the energy-momentum tensor. \\

The main objective of this paper is to investigate the role of $f(Q,T)$ gravity in providing viable cosmological models addressing the late time cosmic speed up issue and any possible role in getting a matter bounce scenario. We emphasize on the mathematical simplification of the extension of the symmetric teleparallel gravity with the change in geometry. In addition, since other geometrically extended theories have been successful in addressing the initial singularity issue, it may be interesting to see whether the gravity based on this combination of teleparallelism and non-metricity would be able to address the same. The paper is organised as: in Sec II a brief review of $f(Q,T)$ gravity and its mathematical formalism have been presented, in Sec III the dynamical parameters of the cosmological  models with matter bounce scenario have been derived. We have considered two different scenarios mimicking a non-singular bounce at some epoch of cosmic evolution. The energy conditions of both the models have been analysed in Sec. IV. To validate the models in the context of geometric dark energy models, the cosmographic test has been performed in Sec. V. The stability analysis of the models are carried out in Sec. VI and in Sec. VII the conclusions are presented. 

\section{A brief review of the $f(Q,T)$ gravity and the Cosmological Taxonomy  }
The action of $f(Q,T)$ gravity \cite{Xu19} is given as,

\begin{equation} \label{eq.1}
S=\int\left(\dfrac{1}{16\pi}f(Q,T)+\mathcal{L}_{m}\right)d^{4}x\sqrt{-g},
\end{equation}

where $\mathcal{L}_{m}$ be the matter Lagrangian and $g=det(g_{\mu \nu})$. The non-metricity $Q$ can be defined as,
\begin{equation} \label{eq.2}
Q\equiv -g^{\mu \nu}( L^k_{~l\mu}L^l_{~\nu k}-L^k_{~lk}L^l_{~\mu \nu}) , 
\end{equation}

where $L^k_{~l\gamma}\equiv -\frac{1}{2}g^{k\lambda}(\bigtriangledown_{\gamma}g_{l\lambda}+\bigtriangledown_{l}g_{\lambda \gamma}-\bigtriangledown_{\lambda}g_{l\gamma})$. Varying the gravitational action \eqref{eq.1}, the field equation of $f(Q,T)$ gravity can be obtained as, 

\begin{equation}\label{eq.3}
-\frac{2}{\sqrt{-g}}\bigtriangledown_{k}(f_{Q}\sqrt{-g}P^{k}_ {\mu \nu})-\frac{1}{2}fg_{\mu \nu}+f_{T}(T_{\mu \nu}+\Theta_{\mu \nu})-f_{Q}(P_{\mu kl} Q^{\;\;\; kl}_{\nu}-2Q^{kl}_{\;\;\;\mu} P_{kl\nu})=8 \pi T_{\mu \nu},
\end{equation}

where $f_Q=\frac{\partial f(Q,T)}{\partial Q}$, the energy momentum tensor, $T_{\mu \nu}\equiv-\frac{2}{\sqrt{-g}}\frac{\delta(\sqrt{-g}\mathcal{L}_m)}{\delta g^{\mu \nu}}$, $\Theta_{\mu \nu}\equiv g^{\alpha \beta}\frac{\delta T_{\alpha \beta}}{\delta g^{\mu \nu}}$.\\

The super potential of the model can be expressed as,
 
\begin{equation} \label{eq.4}
 P^{k}_{\mu \nu}=-\frac{1}{2}L^{k}_{\mu \nu}+\frac{1}{4}(Q^{k}-\tilde{Q}^{k})g_{\mu \nu}-\frac{1}{4}\delta^{k}_{(\mu}Q_{\nu)}.   
\end{equation}

The trace of the energy momentum tensor and trace of the non-metricity tensor can be respectively denoted as,

\begin{eqnarray}
T&=& T_{\mu \nu}g^{\mu \nu}, \nonumber\\
Q_{k}&=& Q_{k}^{\;\;\mu}\;_{\mu}, \tilde{Q}_{k}=Q^{\mu}\;_{k\mu}. \nonumber
\end{eqnarray}\\

We consider the Universe described by homogeneous, isotropic and spatially flat FLRW space-time as,

\begin{eqnarray}\label{eq.5}
ds^{2}=-N^{2}(t)dt^{2}+a^{2}(t)(dx^{2}+dy^{2}+dz^{2}),
\end{eqnarray}
where $N(t)$ and $a(t)$ are respectively be the lapse function and scale factor. The scale factor is related to the Hubble function, which describes the expansion rate as $H(t)=\frac{\dot{a}(t)}{a(t)}$, an over dot represents derivative with respect to cosmic time $t$. The dilation rate is defined as $\tilde{T}=\frac{\dot{N}(t)}{N(t)}$.  The non-metricity for the flat FRW space-time becomes, 
\begin{eqnarray}
 Q &=& 6\frac{H^2}{N^2}.
\end{eqnarray}
In the standard case we have $N(t)=1$ and the non-metricity becomes, $Q=6H^{2}$. We consider a perfect fluid distribution of the Universe and therefore, the energy momentum tensor is written as, $T^{\mu}_{\nu}=diag(-\rho, p, p, p)$. 

Now, the field equations \eqref{eq.3} of $f(Q,T)$ gravity in the standard case for the FLRW space-time can be derived as,
\begin{eqnarray}
p&=&-\frac{1}{16\pi}\left[f-12FH^2-4\dot{\zeta}\right]\label{eq.6} \\
\rho&=&\frac{1}{16\pi}\left[f-12F H^2-4\dot{\zeta}\kappa_{1}\right]\label{eq.7},
\end{eqnarray}
where $F=\frac{\partial f}{\partial Q}$ and $8\pi \kappa\equiv f_{T}=\frac{\partial f}{\partial T}$, $\kappa_{1}=\frac{\kappa}{1+\kappa}$ and $\zeta=FH$. Adding eqns. \eqref{eq.6} and \eqref{eq.7}, the evolution equation of Hubble's function can be obtained as,
\begin{equation} \label{eq.8}
\dot{\zeta}=4\pi(p+\rho)(1+\kappa).
\end{equation}
In comparison to the Friedman equations of Einstein's GR,  the effective pressure ($p_{eff}$) and effective energy density ($\rho_{eff}$) can be characterized as,
\begin{eqnarray}
2\dot{H}+3H^2&=&\frac{1}{F}\left[\frac{f}{4}-2\dot{F}H+4\pi[(1+\kappa)\rho +(2+\kappa)p]\right]=-8\pi p_{eff},\label{eq.9} \\
3H^2&=&\frac{1}{F}\left[\frac{f}{4}-4\pi[(1+\kappa)\rho+\kappa p] \right]=8\pi \rho_{eff}. \label{eq.10}
\end{eqnarray}

Considering cosmological applications, three forms for $f(Q,T)$ have been suggested, such as (i) $f(Q,T)=\lambda_1 Q+\lambda_2 T$, (ii)$f(Q,T)=\lambda_1 Q^{m}+\lambda_2 T$ , (iii)$f(Q,T)=-\lambda_1 Q-\lambda_2 T^2$ \cite{Xu19}. Here $\lambda_1$ and $\lambda_2$ are arbitrary constant. Based on the generality of the forms, we consider here, $f(Q,T)=\lambda_1 Q^{m}+\lambda_2 T$. One can easily get the first model for $m=1$. For this functional, we have $F=\lambda_1 mQ^{m-1}$, $\lambda_2=8\pi\kappa$, $\zeta=\lambda_1mQ^{m-1}H$. Also we have $\dot{F}=2(m-1)F\frac{\dot{H}}{H}$ and $\dot{\zeta}=F\dot{H}(2m-1)$. Now from eqns. \eqref{eq.9} and \eqref{eq.10}, we obtain the pressure and energy density as, 

\begin{eqnarray}\label{eq.11}
p&=& \frac{2\dot{\zeta}[2+\kappa -\kappa \kappa_{1}]-\lambda_1(6H^{2})^{m}(1-2m)}{4\pi [(2+\kappa)(2+3\kappa)-3\kappa^{2}]},  \\
\rho &=&   \frac{2\dot{\zeta}[3\kappa-(2+3\kappa)\kappa_{1}]+\lambda_1 (6H^2)^{m} (1-2m)}{4\pi[(2+\kappa)(2+3\kappa)-3\kappa^{2}]}. \label{eq.12}
\end{eqnarray}
The equation of state (EoS) parameter $\omega=\frac{p}{\rho}$ can be obtained as,
\begin{equation}
\omega =\frac{p}{\rho } = -1+\frac{4\dot{\zeta}[(1+2\kappa )(1-\kappa_{1})]}{2\dot{\zeta}[3\kappa-(2+3\kappa)\kappa_{1}]+\lambda_1 (6H^2)^{m} (1-2m)}   \label{eq.13}
\end{equation}
 
The violation of strong energy condition is essential for the validity of extended  theory of gravity, so we discuss here the energy conditions. We may express the general statement of the energy conditions and their forms for the energy-momentum tensor  \cite{Novello08}, \\

$$T^{\mu}_{\nu}=diag(-\rho, p, p, p)$$.\\

For each null vector, the null energy condition (NEC) asserts that\\

$NEC\iff T_{\mu\nu}k^{\mu}k^{\nu}\geq 0$ $\Rightarrow$ $\rho +p\geq 0$.\\

For any timelike vector, the weak energy condition (WEC) asserts that\\

$WEC\iff T_{\mu\nu}v^{\mu}v^{\nu}\geq 0$ $\Rightarrow$ $ \rho\geq 0 \ \text{and} \ \rho +p\geq 0$.\\

For any timelike vector, the strong energy condition (SEC) asserts that\\

$SEC\iff\left(T_{\mu\nu}-\frac{T}{2}g_{\mu\nu}\right)v^{\mu\nu}\geq 0$ $\Rightarrow$ $ \rho +3p\geq 0$.\\

For any timelike vector, the dominant energy condition (DEC) asserts that\\

$DEC\iff T_{\mu\nu}v^{\mu}v^{\nu}\geq 0$ $\Rightarrow$ $\rho -p\geq 0$ and $T_{\mu\nu}v^{\mu}$ is not spacelike.\\

Moreover, it has been observed that if null energy condition is violated then all the other pointwise energy conditions would be violated. However, for dark energy dominated models with negative pressure cosmic fluid, the DEC is satisfied even if the NEC is violated. In some bouncing models or models dominated by phantom fields, the NEC has to be violated. Now, we can express the general form of energy conditions in the context of $f(Q,T)=\lambda_1Q^m+\lambda_2T$ gravity as,
\begin{eqnarray}
\rho+p&=& \frac{1}{4\pi}\left[(1-\kappa_1)\dot{\zeta}\right], \label{eq.13a}\\
\rho+3p&=&\frac{1}{16\pi}\left[-2f+24FH^2+4\dot{\zeta}(3-\kappa_1)\right] \nonumber \\
&=& \frac{1}{16 \pi(1+2\kappa)}\left[-2(1-2m)\lambda_1Q^m+2\dot{\zeta}(6+6\kappa-2\kappa_1-6\kappa\kappa_1)\right], \label{eq.13b}\\
\rho-p&=&\frac{1}{8\pi}\left[f-12FH^2-2\dot{\zeta}(1+\kappa_1)\right] \nonumber \\
&=&\frac{1}{8\pi(1+2\kappa)}\left[(1-2m)\lambda_1Q^m+2\dot{\zeta}(-1+\kappa-\kappa_1-\kappa\kappa_1)\right].
\end{eqnarray}

In the present work, we are interested to discuss some bouncing models and therefore we expect that the NEC should be violated through out the cosmic evolution. The details of the evolutionary behaviour of the energy conditions will be discussed in the respective models. However, we wish to present the conditions necessary for NEC violation within the framework of $f(Q,T)$ gravity. 
Eq.\eqref{eq.13a} ensures that a violation of NEC i.e $\rho+p<0$ leads to the condition that either $\kappa_1<1$ or we have a negative value for $\dot{\zeta}$. If the first condition holds, then we should have $\kappa=\frac{1}{8 \pi}\frac{\partial f}{\partial T} <-1$. On the other hand, if the second condition holds, we have $\dot{\zeta}<0$ which leads to $\dot{H}\frac{\partial f}{\partial Q}+\frac{\partial}{\partial t}\left(\frac{\partial f}{\partial Q}\right)H <0$. All the symmetric bounce models satisfy the conditions $H=0$ and $\dot{H}>0$ at  the bouncing epoch. In view of this, at least at the bouncing epoch, a violation of NEC with $\kappa_1>0$ requires $\frac{\partial f}{\partial Q}<0$. In an earlier work \cite{Pati21}, we have shown that, $f(Q,T)$ is a negative quantity. Also, it is a decreasing function of the redshift. In view of this, the second condition is mostly satisfied.

\section{Dynamical parameters of some Cosmological Models favouring Matter Bounce Scenario}

As an application of the formalism developed in the previous section, we wish to consider some matter bounce scenario, where the Universe is assumed to contract before the non-singular bouncing epoch followed by a matter dominated expansion. Such a scenario has been proposed to avoid the singularity occurring in the phenomenal Big Bang models. In view of this, in the following subsections, we will consider two different matter bounce scenario that are motivated from loop quantum gravity. The quantum signature of the model is handled through certain model parameters which will be tuned so as to get viable bouncing models that also provide suitable explanation for the late time cosmic speed up issue.
\subsection{Model I}
We consider the ansatz for the scale factor as,
\begin{equation}
a(t)=\left(\frac{\alpha}{\chi}+t^2 \right)^{\frac{1}{2\chi}},
\end{equation} 
where $\alpha$ and $\chi$ are scale factor parameters which are chosen suitably to provide a bouncing behaviour. This model bounces at $t_b=0$ with a non zero finite value for the scale factor i.e $a(t_b)=\left(\frac{\alpha}{\chi}\right)^{\frac{1}{2\chi}}$. Before this bouncing epoch, the scale factor expands symmetrically in both sides of the bounce point giving us a picture of early contraction followed by a bounce and then an expansion. As a new era arises, the scale factor experiences fluctuations as it undergoes a transition from a contracting to an expanding stage. In the Big Bang cosmological norm, this stage transition results in a non-singular bounce. The Hubble parameter for this scale factor may be written as

\begin{equation}
H={t}(\alpha +t^2 \chi )^{-1}, 
\end{equation}
and in terms of the redshift $z=-1+\frac{a_0}{a}$ as 
\begin{equation} \label{eq.14}
 H(z)=\frac{\left({\chi -\alpha (1+z)^{2\chi}}\right)^{\frac{1}{2}}(1+z)^{\chi}}{\chi^{\frac{3}{2}}},  
\end{equation}
where  $a_0$ is the scale factor at the present epoch. The Hubble parameter changes from $H(t)<0$ to $H(t)>0$ through $H=0$ at the bounce point. The evolutionary behaviour of the Hubble parameter for the above bouncing scenario is shown in FIG.1. In order to plot the figure, we have used the parametric values of the constants as $\alpha=0.43$ and $\chi=1.001$. The Hubble parameter satisfies the prescribed conditions for the bouncing cosmology as it comes from a negative value, passing through zero at $t=0$, and then it has a positive behaviour. In the figure, we have also shown the Hubble parameter as a function of the redshift for the positive time zone only.  As a function of redshift, the Hubble parameter is found to increase  sharply at a past epoch and after the initial epoch, it decreases linearly into the future. 

\begin{figure}[tbph]
\minipage{0.50\textwidth}
\centering
\includegraphics[width=\textwidth]{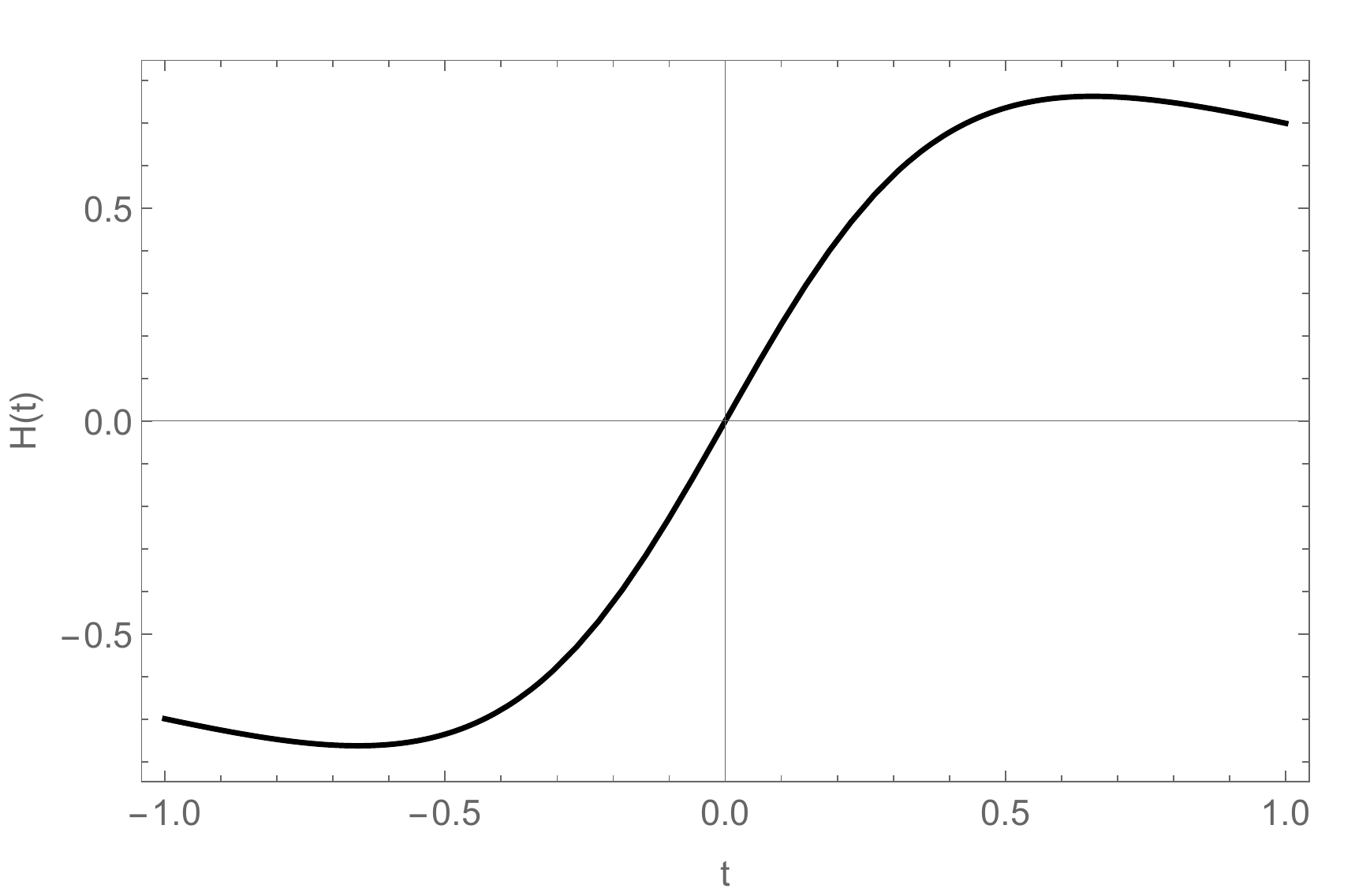}
\endminipage\hfill
\minipage{0.50\textwidth}
\includegraphics[width=\textwidth]{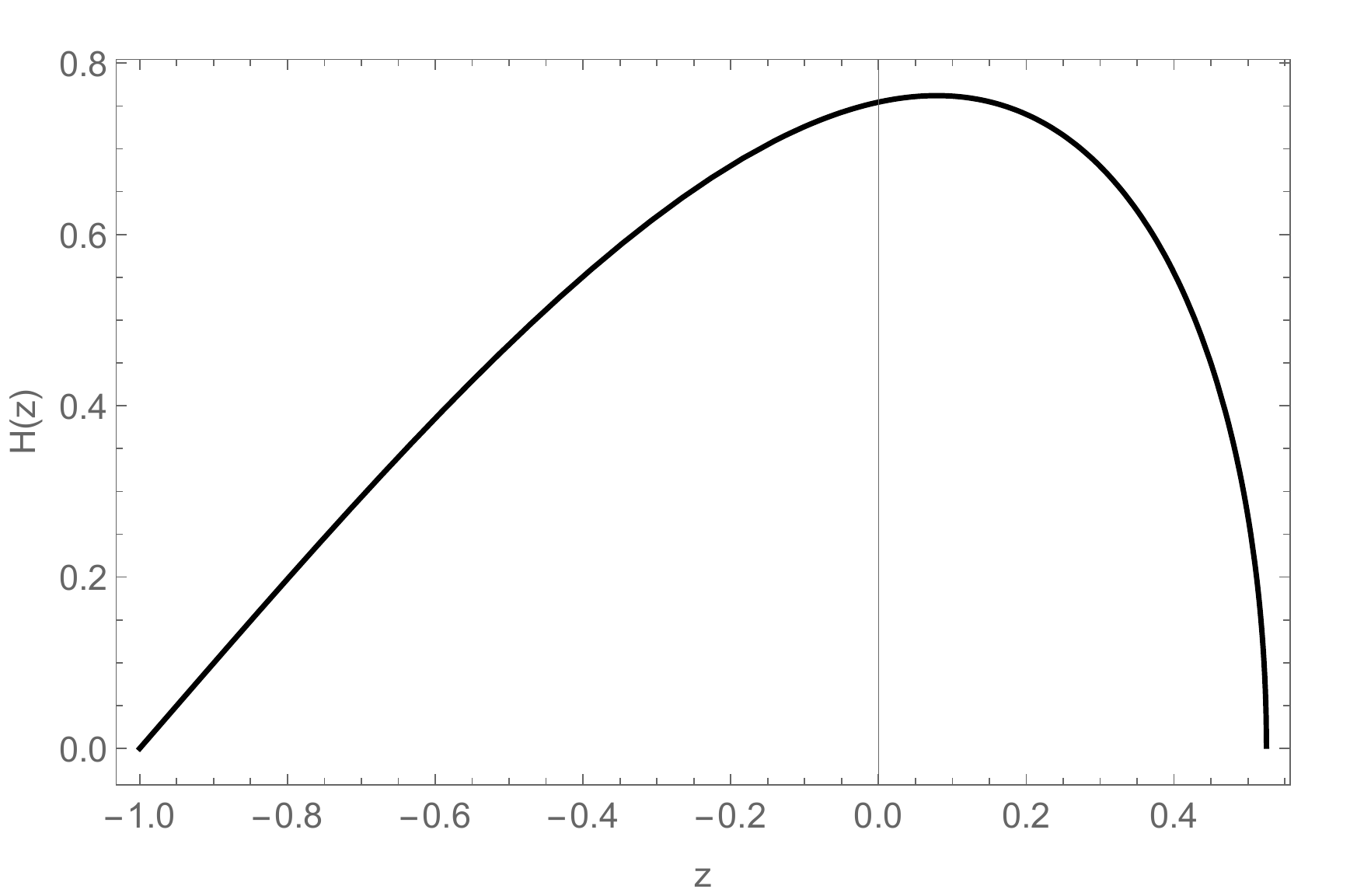} 
\endminipage
\caption{Plot for the variation of the Hubble parameter vs. cosmic time (Left panel) and Hubble parameter vs. redshift (Right panel) for Model I with $\alpha=0.43$ and $\chi=1.001$~~~~[Eq. \eqref{eq.14}]}
\end{figure}
It is important to study the dynamical parameters of the model to understand the dynamical aspects of the Universe. The behaviour of dynamical parameters gives a quick indication of whether or not the model is bouncing. The dynamical parameters such as the pressure, energy density, and the EoS parameter for the modified symmetric teleparallel gravity theory $f(Q, T)$ can be easily derived from equations \eqref{eq.11}-\eqref{eq.13}  and are represented as \eqref{eq.15}-\eqref{eq.17} where we have used the shorthand notation $\tilde{\lambda}_2=\lambda_2+8\pi$.
\begin{equation}\label{eq.15}
p=\frac{2^{m-2}3^{m-1}(1-2m)(1+z)^{2m\chi}\lambda_1\left[\chi(\left(3\tilde{\lambda}_2-8\pi\right)m\chi-3\tilde{\lambda}_2)+\alpha \left(3\tilde{\lambda}_2(1-2m \chi)+16\pi m \chi\right)(1+z)^{2\chi}\right]}{\chi^{3m}\tilde{\lambda}_2(\tilde{\lambda}_2-4\pi)\left[\chi-\alpha(1+z)^{2\chi}\right]^{1-m}}, 
\end{equation}
\begin{equation}\label{eq.16}
\rho = \frac{2^{m-2}3^{m-1}(1-2m)(1+z)^{2m\chi}\lambda_1\left[\chi(\lambda_2m\chi+3\tilde{\lambda}_2)-\alpha(2\lambda_2m\chi+3\tilde{\lambda}_2)(1+z)^{2\chi}\right]}{\chi^{3m}\tilde{\lambda}_2(\tilde{\lambda}_2-4\pi)\left[\chi-\alpha(1+z)^{2\chi}\right]^{1-m}},
\end{equation}
\begin{equation}\label{eq.17}
\omega =-1+4m\chi\left(\tilde{\lambda}_2-4\pi\right)\frac{\chi-2\alpha(1+z)^{2\chi}}{\chi(\lambda_2m\chi+3\tilde{\lambda}_2)-\alpha(2\lambda_2m\chi+3\tilde{\lambda}_2)(1+z)^{2\chi}}.
\end{equation}
\begin{figure}[tbph]
\minipage{0.50\textwidth}
\centering
\includegraphics[width=\textwidth]{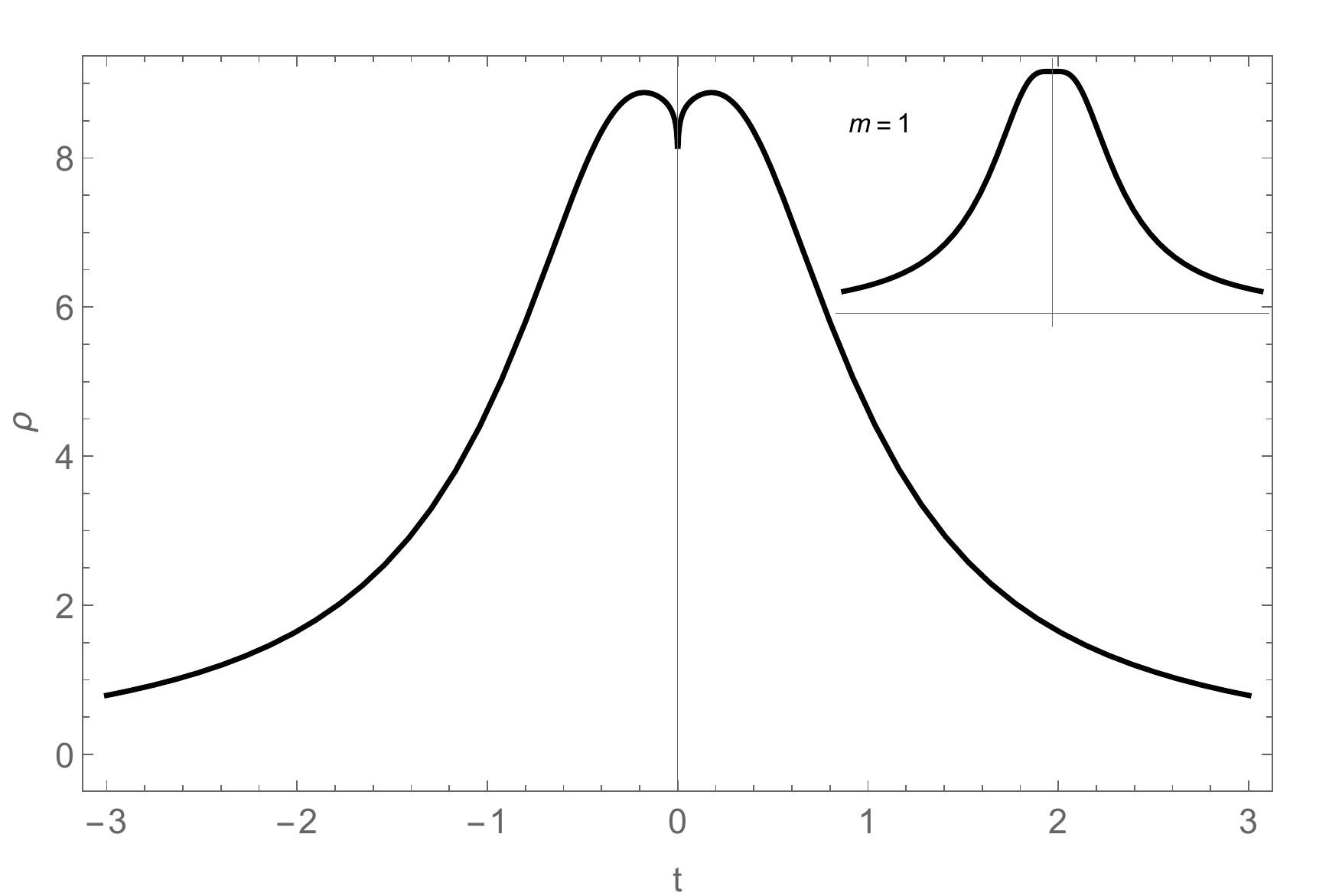}
\endminipage\hfill
\minipage{0.50\textwidth}
\includegraphics[width=\textwidth]{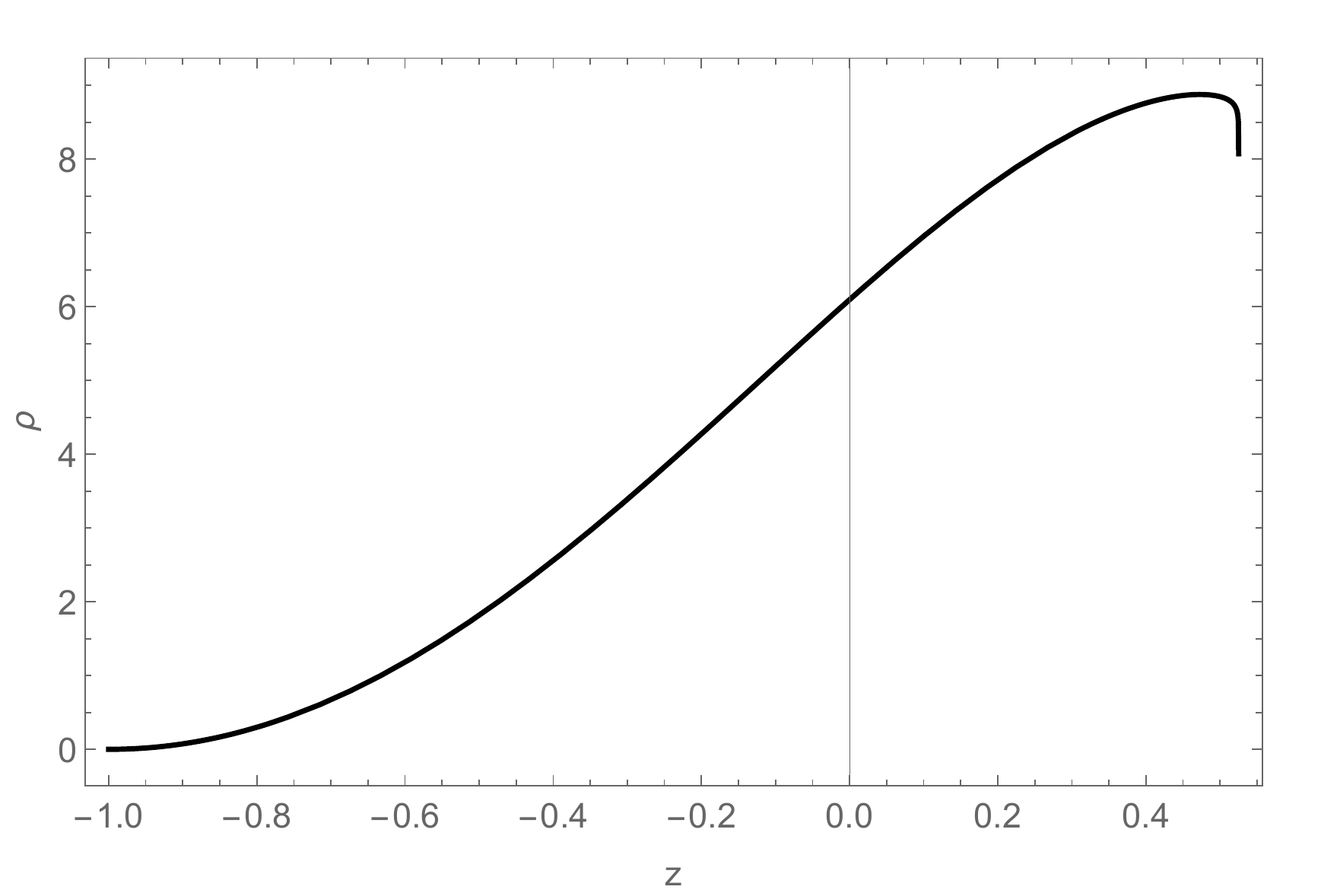} 
\endminipage
\caption{Plot for the variation of the energy density vs. cosmic time (Left panel) and the energy density vs. redshift (Right panel) for Model I. We have used the parameter space $\alpha=0.43$, $\chi=1.001$, $\lambda_1=-0.5$, $\lambda_2=-12.5$ and $m=1.01$.~~[Eq. \eqref{eq.15}]}
\minipage{0.50\textwidth}
\centering
\includegraphics[width=\textwidth]{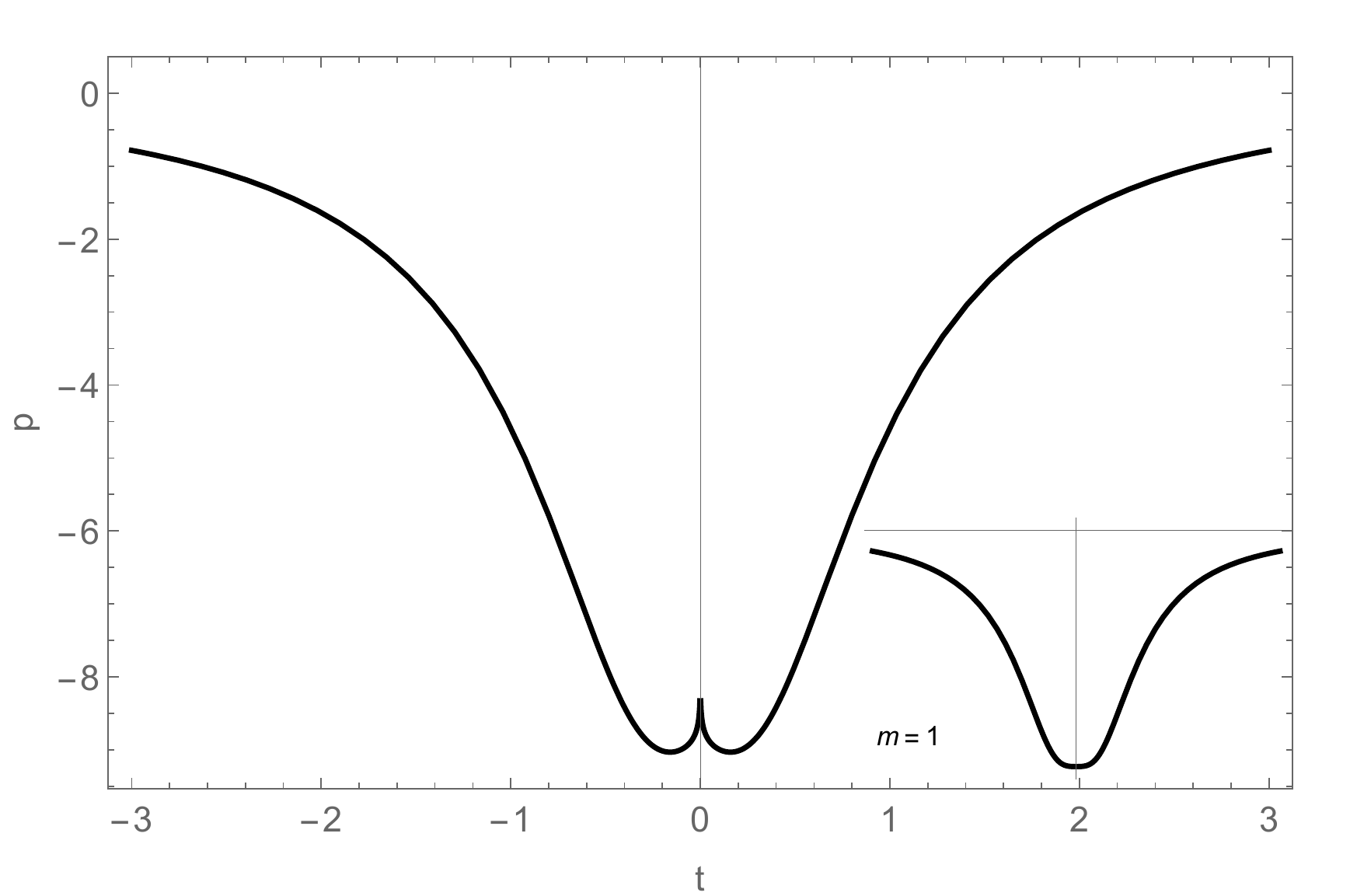}
\endminipage\hfill
\minipage{0.50\textwidth}
\includegraphics[width=\textwidth]{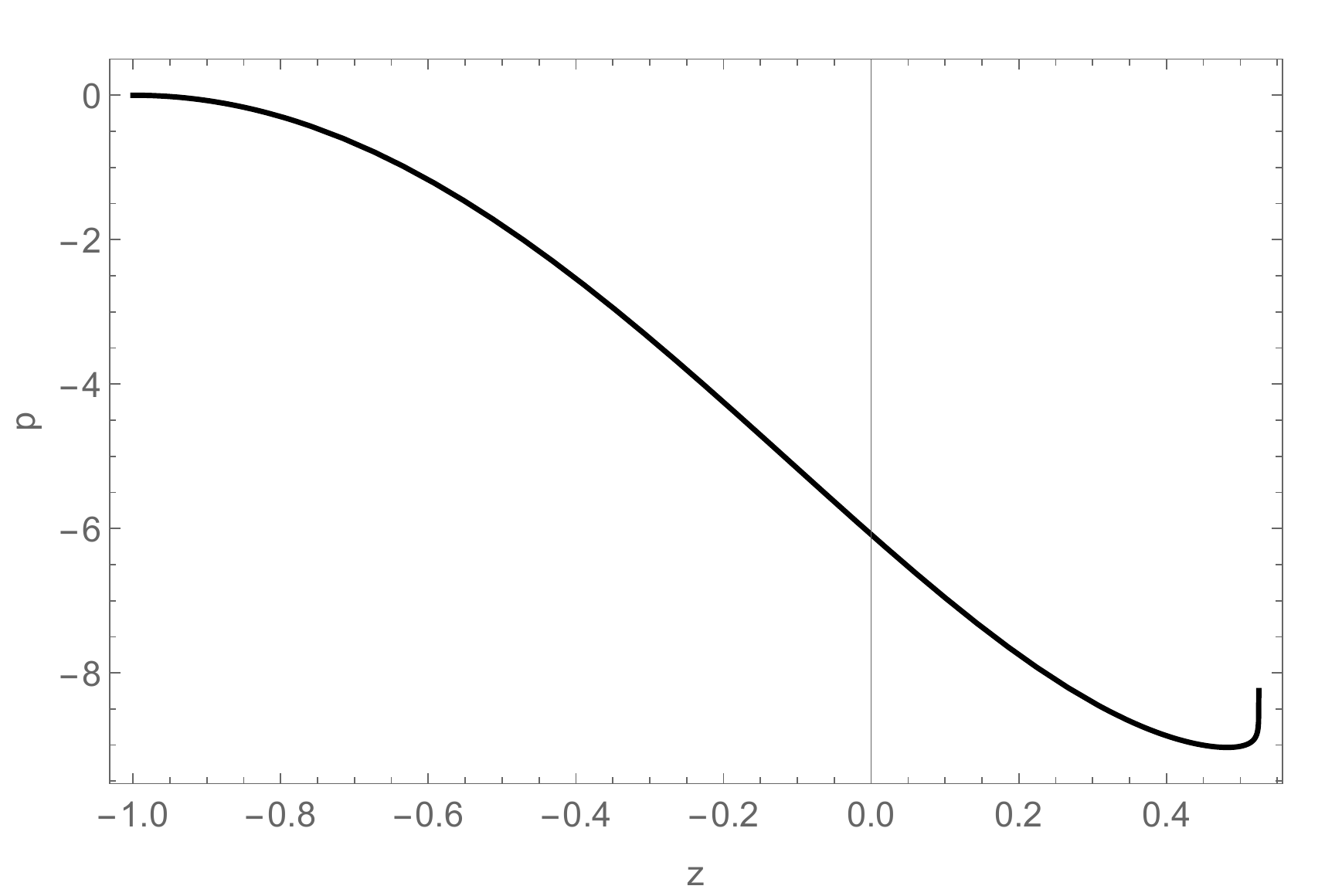} 
\endminipage
\caption{Plot for the variation of the pressure vs. cosmic time (Left panel) and the pressure vs. redshift (Right panel) for Model I with the $\alpha=0.43$, $\chi=1.001$, $\lambda_1=-0.5$, $\lambda_2=-12.5$ and $m=1.01$.~~[Eq. \eqref{eq.16}]}
\minipage{0.50\textwidth}
\centering
\includegraphics[width=\textwidth]{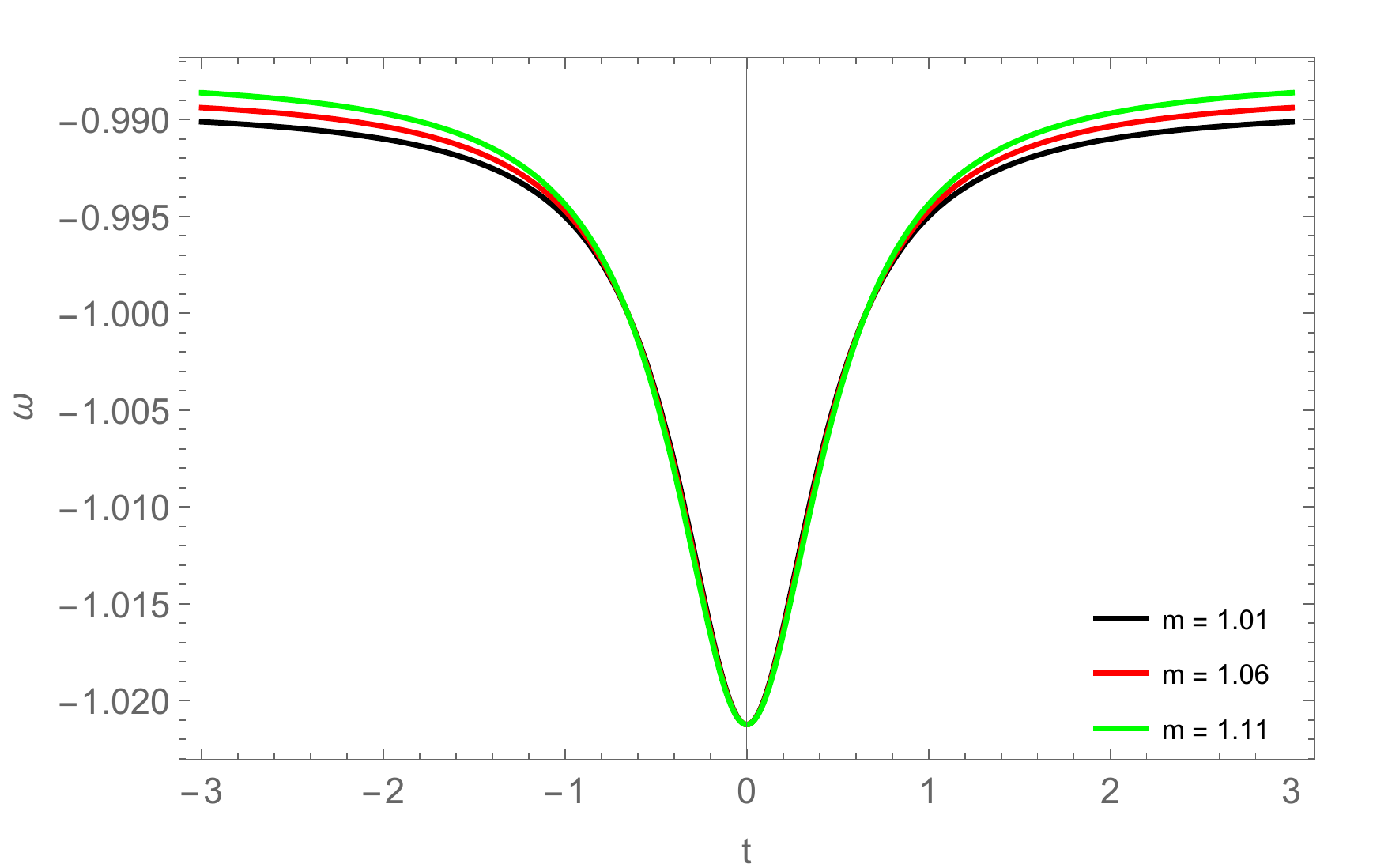}
\endminipage\hfill
\minipage{0.50\textwidth}
\includegraphics[width=\textwidth]{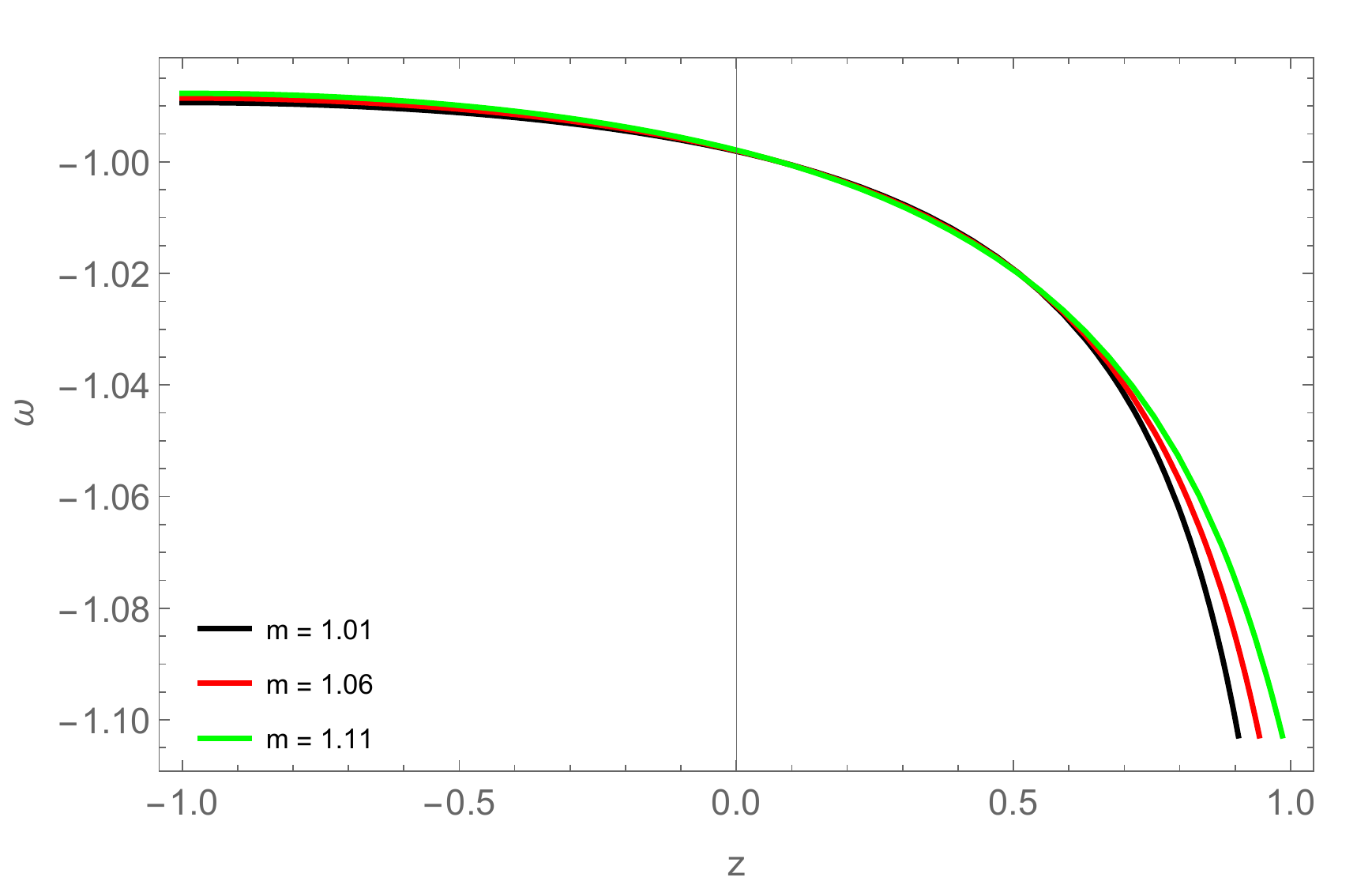} 
\endminipage
\caption{Plot for the variation of the EoS parameter vs. cosmic time (Left panel) and the EoS parameter vs. redshift (Right panel)  for Model I with the $\alpha=0.43$, $\chi=1.001$, $\lambda_1=-0.5$, $\lambda_2=-12.5$ and $m=1.01, 1.06, 1.11$.~~[Eq. \eqref{eq.17}]}
\end{figure}\\

The graphical representation for the energy density is shown in FIG.2 both as function of cosmic time (left panel) and redshift (right panel). In order to ensure a positive energy density through out the cosmic evolution including the negative time zone, we have considered $\lambda_1=-0.5$, $\lambda_2=-12.5$ and $m=1.01$. For this choice of the parameters, the energy density is observed to increase in the pre bounce epoch as we move closer to the bouncing point. It forms a ditch near the bounce and then in the post bounce region, it decreases. For a choice of the teleparallel gravity parameter $m=1$ (inset in the left panel figure), the ditch in the energy density near the bounce disappears. With respect to the redshift (right panel of the figure), the energy density decreases from a higher positive value to a smaller positive value. The graphical representation of pressure with cosmic time and redshift is shown in FIG.3. The graphical representation for pressure looks like a mirror image of the curve for energy density. The pressure is a negative quantity through out the cosmic evolution. In the pre bounce region, it decreases from a small negative value to large negative value at the bounce. On the other hand, in the post bounce region, it increases from large negative values to small negative values. A small bump is formed in the curve of the pressure at the bouncing epoch which disappears for $m=1$. In view of this, we believe that, the choice of the model parameter $m$ has a substantial effect on the dynamical aspects of the model. In the right panel of FIG.3, the pressure is shown as a function of redshift. One may observe that, the pressure increases almost linearly after showing a sharp decrease in an initial epoch around $z=0.525$. The $p<0$ and $\rho>0$ behaviour in the post bounce period, illustrates an accelerating Universe.

In FIG.4, we show the evolution of the EoS parameter both as function of cosmic time (left panel) and redshift ( right panel). As we move from the pre bounce period to post bounce period through the bouncing epoch, the EoS parameter is observed to decrease initially, passes the phantom divide and then increases again after forming a well near the bounce. The depth of the well depends on the choice of the parameter $m$. The behaviour of the EoS parameter in the right panel shows that in the post bounce period, it increases from a phantom field dominated phase to a quintessence like phase. The role of the modified gravity parameter is to split the tail region of the EoS parameter because the rate of increment in $\omega$ is higher for large values of $m$. For all the values of $m$ chosen in the present work,  $\omega$ tends to $-0.99$ as $z$ tends to -1, and at $z=0$ it has a value close to $-0.998$. This prediction shows that, the present model provides viable bouncing scenario alongside providing a suitable explanation for the late time cosmic speed up issue.

\subsection{Model II}
As a second example, we consider the ansatz,
\begin{equation}\label{eq.22a}
a(t)=\sqrt[3]{\frac{3 \rho_{c}t^2}{4}+1},
\end{equation}
where $\rho_c$ is a constant parameter for the scale factor $a(t)$. The model bounces at the epoch $t_b=0$ and the scale factor at bounce is $a (t_b)=1$. The Hubble parameter for the scale factor becomes, 
\begin{equation}\label{eq.22b}
H=\frac{\rho_{c}t}{2}\left(\frac{3\rho_{c}t^2} {4}+1\right)^{-1}.
\end{equation}

In terms of redshift, the Hubble parameter for the present model can be expressed as,

\begin{equation}\label{eq.22}
H(z)=\frac{\sqrt{\rho_{c}}}{\sqrt{3}} \left(1- (1+z)^{3}\right)^{\frac{1}{2}}(1+z)^{\frac{3}{2}}.
\end{equation}

In FIG. 5, the Hubble parameter is shown as a function of cosmic time (left panel) and as a function of redshift (right panel). We chose the parameter $\rho_c = 0.75$. The Hubble parameter satisfies the required conditions of bouncing cosmology. The Hubble parameter exhibits negative behaviour in the early stages of evolution, passes through zero at $t=0$, and then exhibits positive behaviour in the latter stages of evolution. In  other words, $H<0$ in the pre bounce phase, $H=0$ at bounce and $H>0$ during the post bounce epoch.  Also, we have $\dot{H}>0$ at bounce.  The behaviour of the Hubble parameter with respect to the redshift is interesting. It has a sharp increase in value initially and it reaches its maximum value at $z=-0.2$. After that, it starts to decrease almost linearly and as $z$ approaches $-1$, the Hubble parameter falls to zero. 

\begin{figure}[tbph]
\minipage{0.50\textwidth}
\centering
\includegraphics[width=\textwidth]{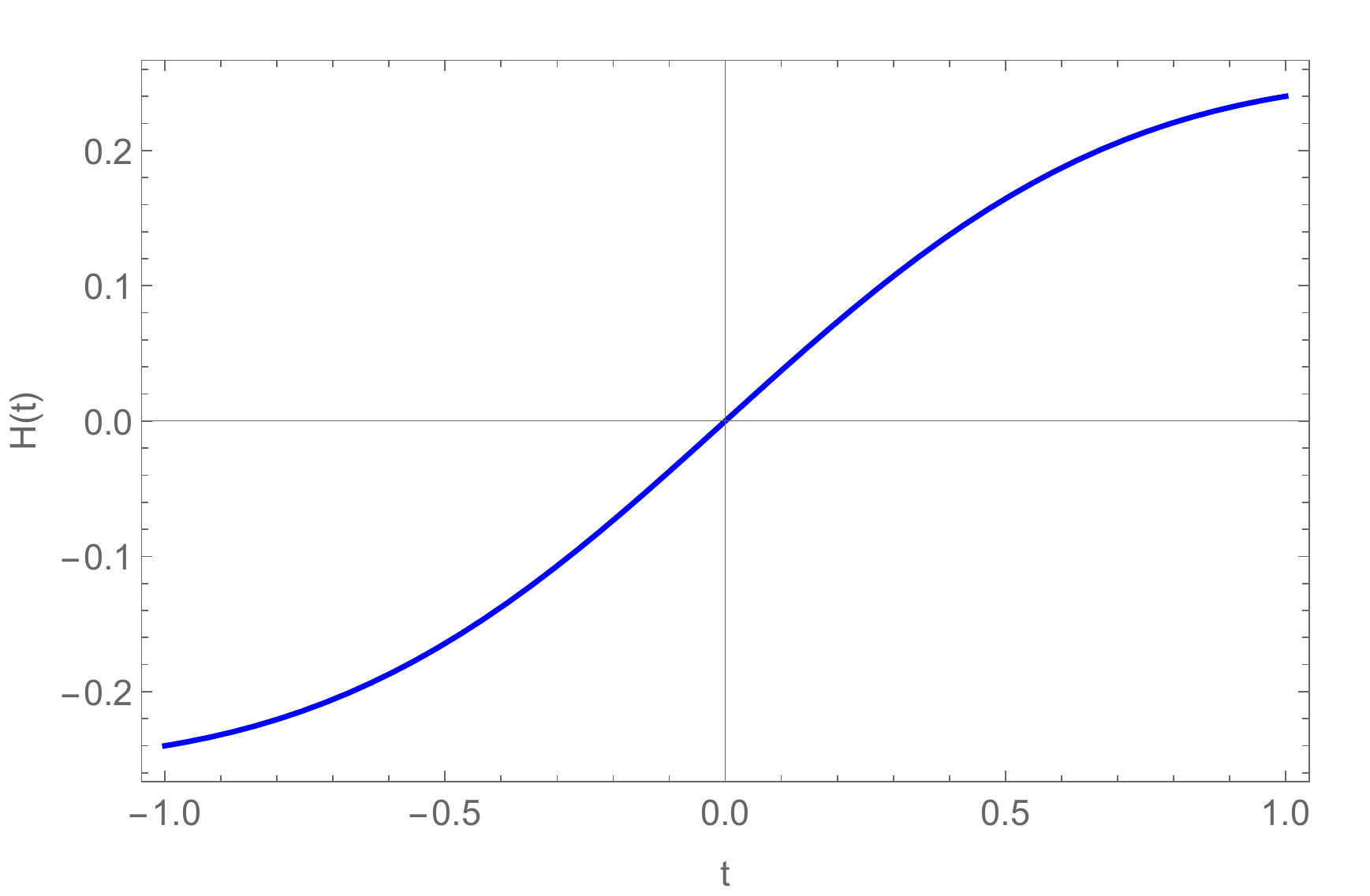}
\endminipage\hfill
\minipage{0.50\textwidth}
\includegraphics[width=\textwidth]{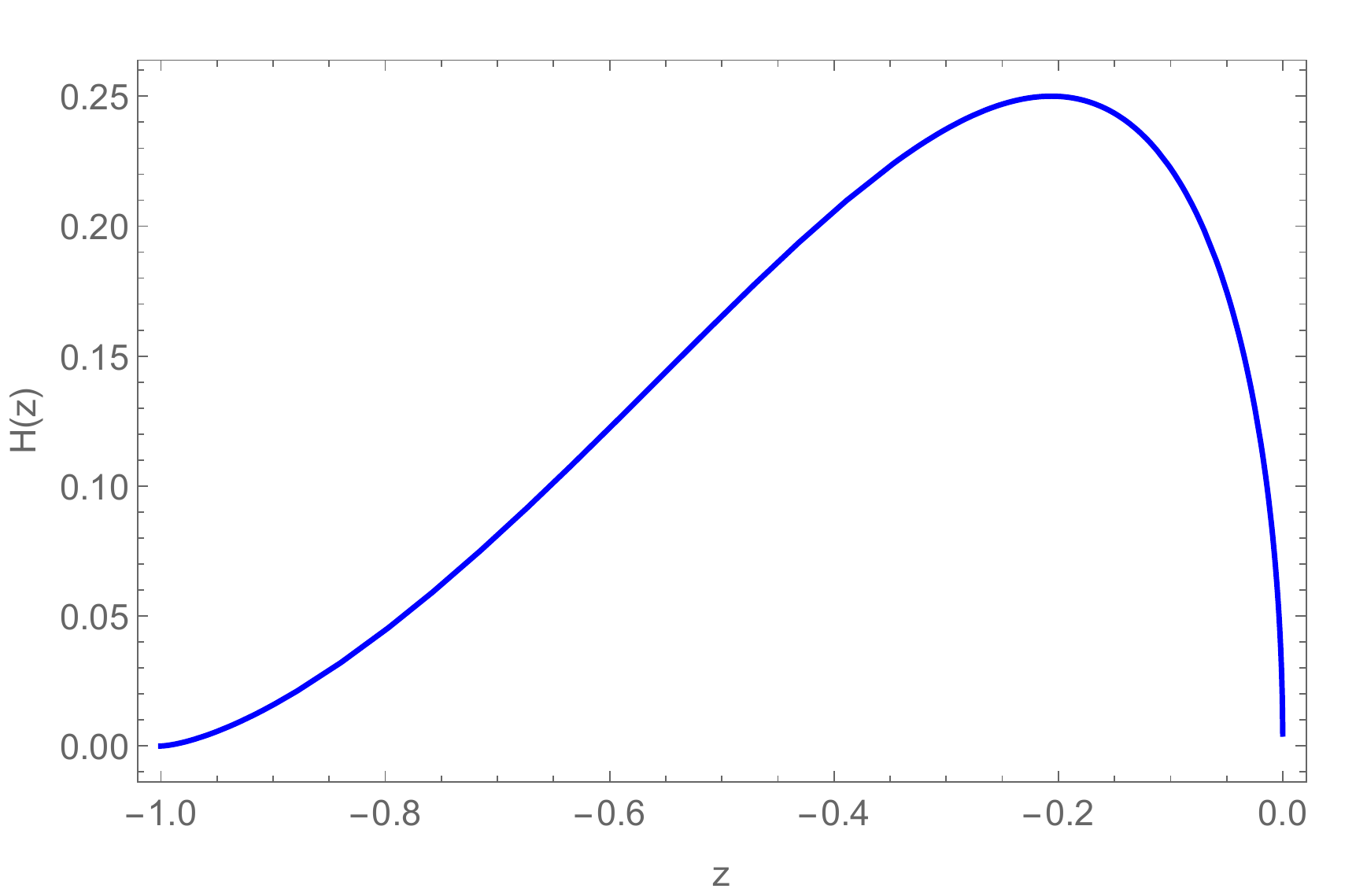} 
\endminipage
\caption{Plot for the variation of the Hubble parameter vs. cosmic time (Left panel) and Hubble parameter vs. redshift (Right panel) for Model II with $\rho_{c}=0.75$.~~[Eq.\eqref{eq.22}]]}
\end{figure}
The pressure, energy density and the EoS parameter  for the present model are obtained from the Eqs. \eqref{eq.11}- \eqref{eq.13} and the scale factor in Eq.\eqref{eq.22a} as
\begin{equation}\label{eq.23}
p=\frac{2^{m-3}(2m-1)\rho_{c}^{m}\lambda_1\left[2z\tilde{\lambda}_2(z^{2}+3z+3)-m(3\tilde{\lambda}_2-8\pi)(2z^{3}+6z^{2}+6z+1)\right]}{\tilde{\lambda}_2\left(\tilde{\lambda}_2-4\pi\right)z(z^{2}+3z+3)[(1+z)^{3}-(1+z)^{6}]^{-m}},
\end{equation}
\begin{equation}\label{eq.24}
\rho=-\frac{2^{m-3}(2m-1)\rho_{c}^{m}\lambda_1\left[2z(\lambda_2m+\tilde{\lambda}_2)(z^{2}+3z+3)+\lambda_2m\right]}{\tilde{\lambda}_2\left(\tilde{\lambda}_2-4\pi\right)(z^{3}+3z^{2}+3z)[(1+z)^{3}-(1+z)^{6}]^{-m}},
\end{equation}
\begin{equation} \label{eq.25}
\omega=-1+\frac{2[\lambda_2m+2\tilde{\lambda}_2](z^{3}+3z^{2}+3z)-m(3\tilde{\lambda}_2-8\pi)(2z^{3}+6z^{2}+6z+1)+\lambda_2m}{2[\lambda_2m+\tilde{\lambda}_2](z^{3}+3z^{2}+3z)+\lambda_2m}. \end{equation}

FIG. 6  depicts the graphical behaviour of the energy density as a function of cosmic time and redshift. The energy density for the typical values of the model parameters demonstrates the bounce at $t=0$ with a well-shaped curve. The parameters have been limited to ensure that the energy density remains positive during the bounce and in both the negative and positive time zones. The behaviour of the energy density is similar to the previous bouncing model discussed in Model I. It evolves from a small  positive value to form a ditch at bounce and then it decreases in the positive time zone. The peak value of  the energy density and the formation of the ditch depends on the choice of the parameter $m$. For $m=1$, we do not have such ditch in the curve for the energy density and the energy density curve almost follow a Gaussian pattern. The pressure for the present model is shown in FIG. 7. The pressure evolves with negative values through out. In the negative time zone, as the cosmic time approaches the bouncing epoch, the value of the pressure drops more rapidly. On the other hand, in the post bounce period, the pressure increases rapidly after forming a bump near the bounce. The formation of the bump depends on the choice of $m$. As usual, for $m=1$, the bump in the pressure curve disappears.

We have presented the graphical behaviour of the EoS parameter with cosmic time and redshift parameter in FIG. 8.  Throughout the cosmic evolution, the EoS parameter remains near the concordant $\Lambda$CDM  value $\omega=-1$.  However, if looked closely to the dynamical aspect of the EoS parameter, it evolves from a lower negative value, crosses the phantom divide and attains a minimum in the phantom like region at the bouncing epoch. In the post bounce period, it again rises from a phantom like region to a quintessence region. At the present epoch, it assumes a value of $\omega_0=-1.002$ which is closer to some recent estimates.
\begin{figure}[tbph]
\minipage{0.50\textwidth}
\centering
\includegraphics[width=\textwidth]{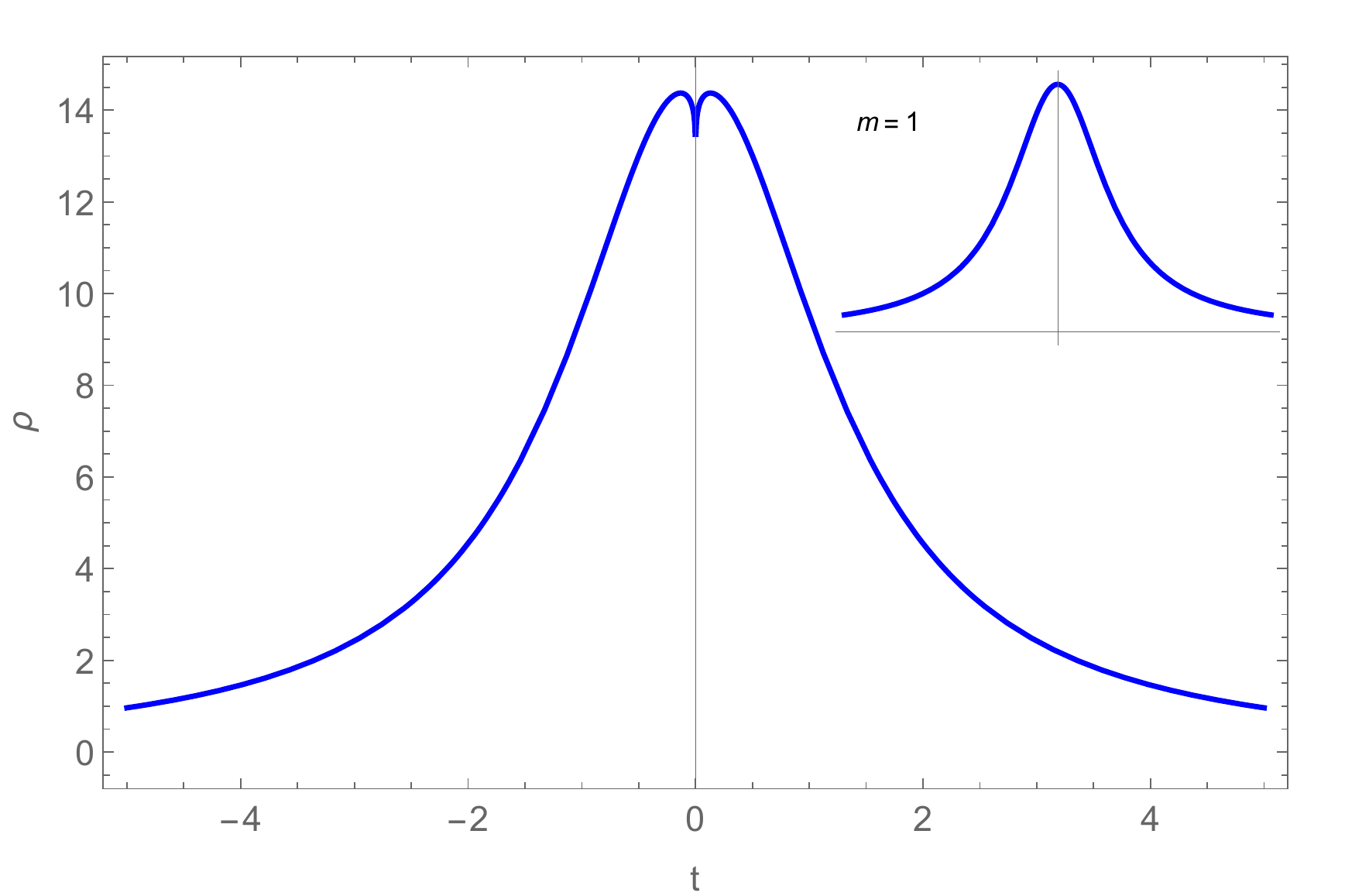}
\endminipage\hfill
\minipage{0.50\textwidth}
\includegraphics[width=\textwidth]{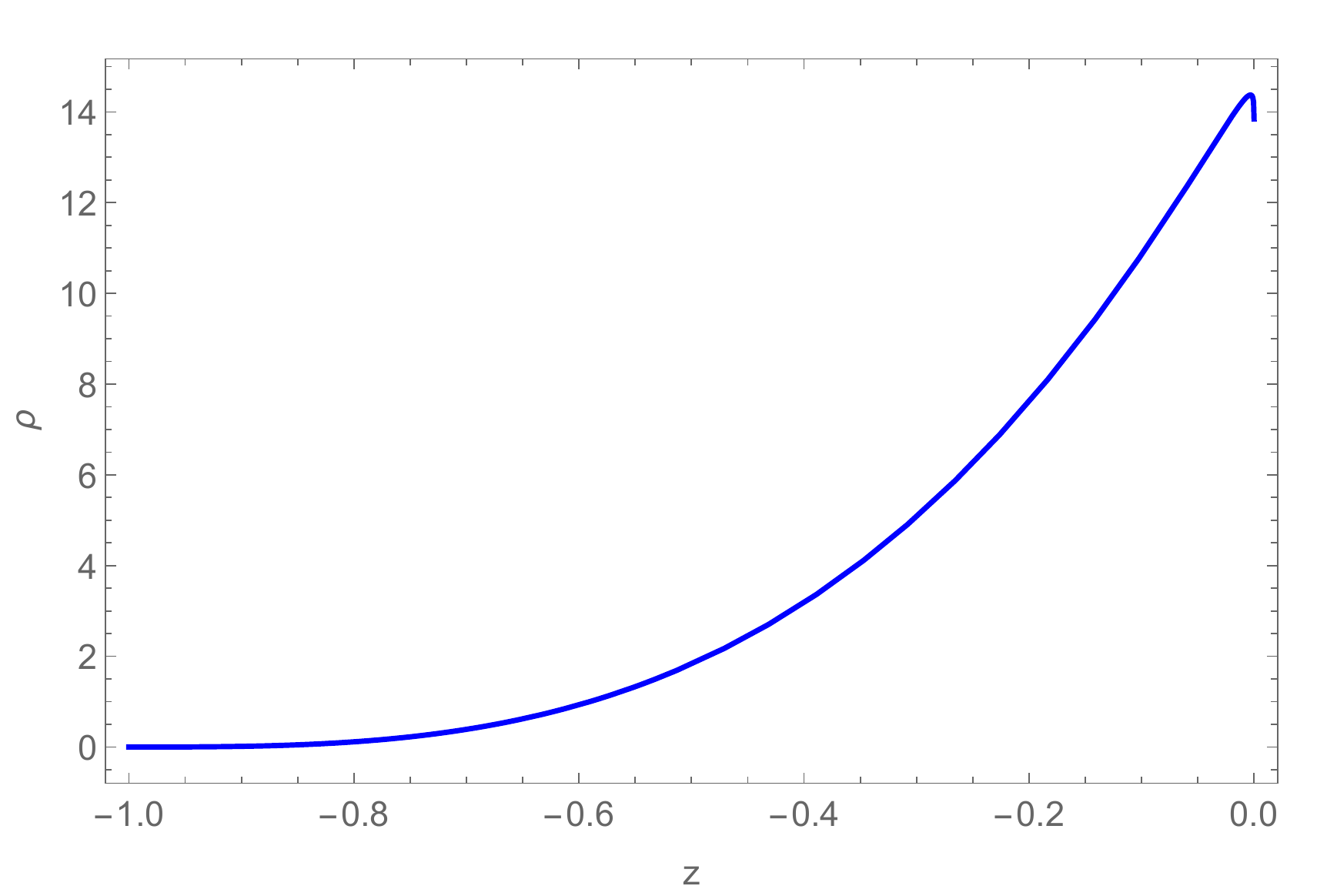}
\endminipage
\caption{Plot for the variation of the energy density vs. cosmic time (Left panel) and  energy density vs. redshift (Right panel)for Model II with the $\rho_{c}=0.75$, $\lambda_1=-0.5$, $\lambda_2=-12.56$ and $m=1.01$.~~[Eq.\eqref{eq.23}]}
\minipage{0.50\textwidth}
\centering
\includegraphics[width=\textwidth]{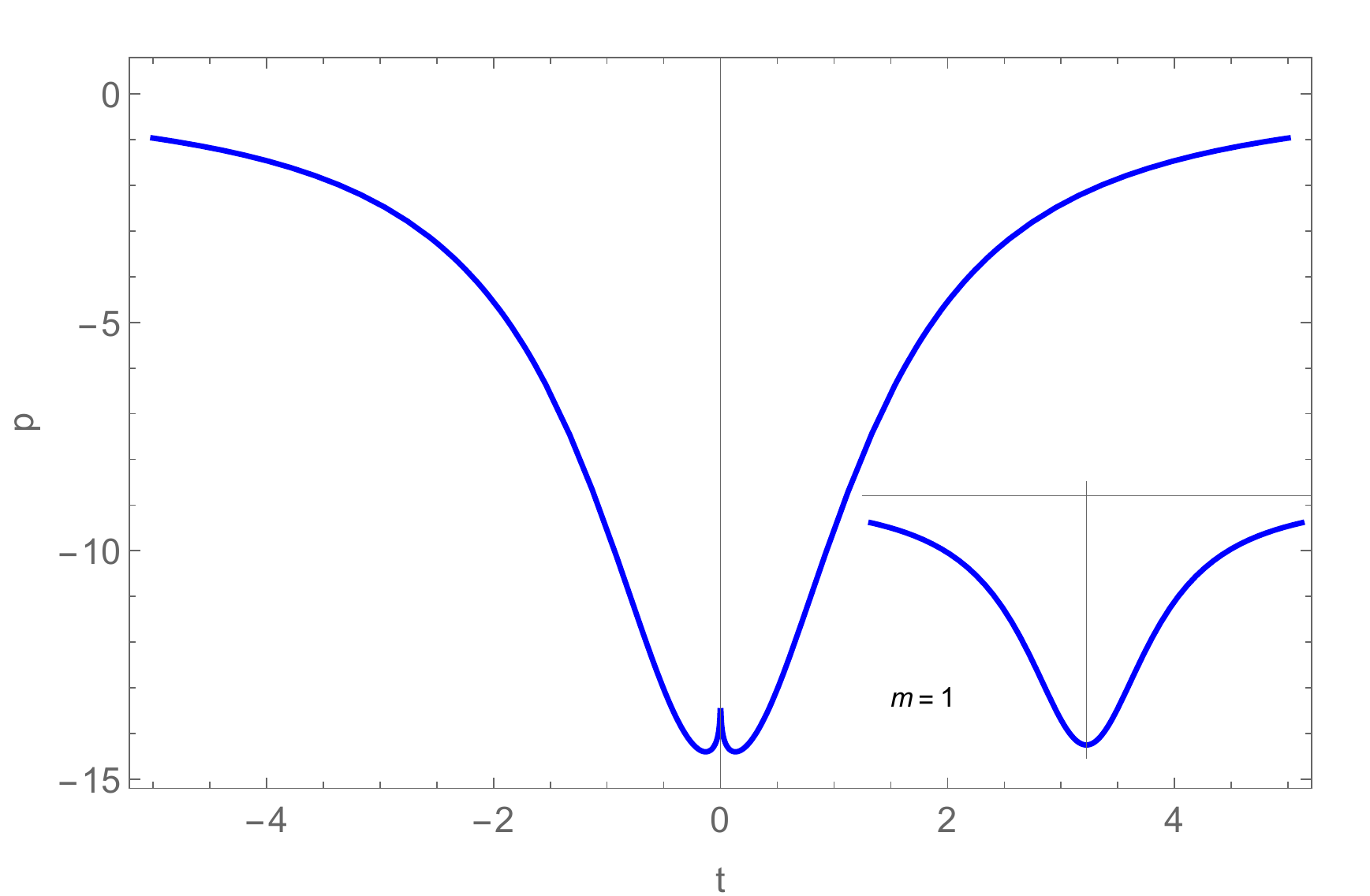}
\endminipage\hfill
\minipage{0.50\textwidth}
\includegraphics[width=\textwidth]{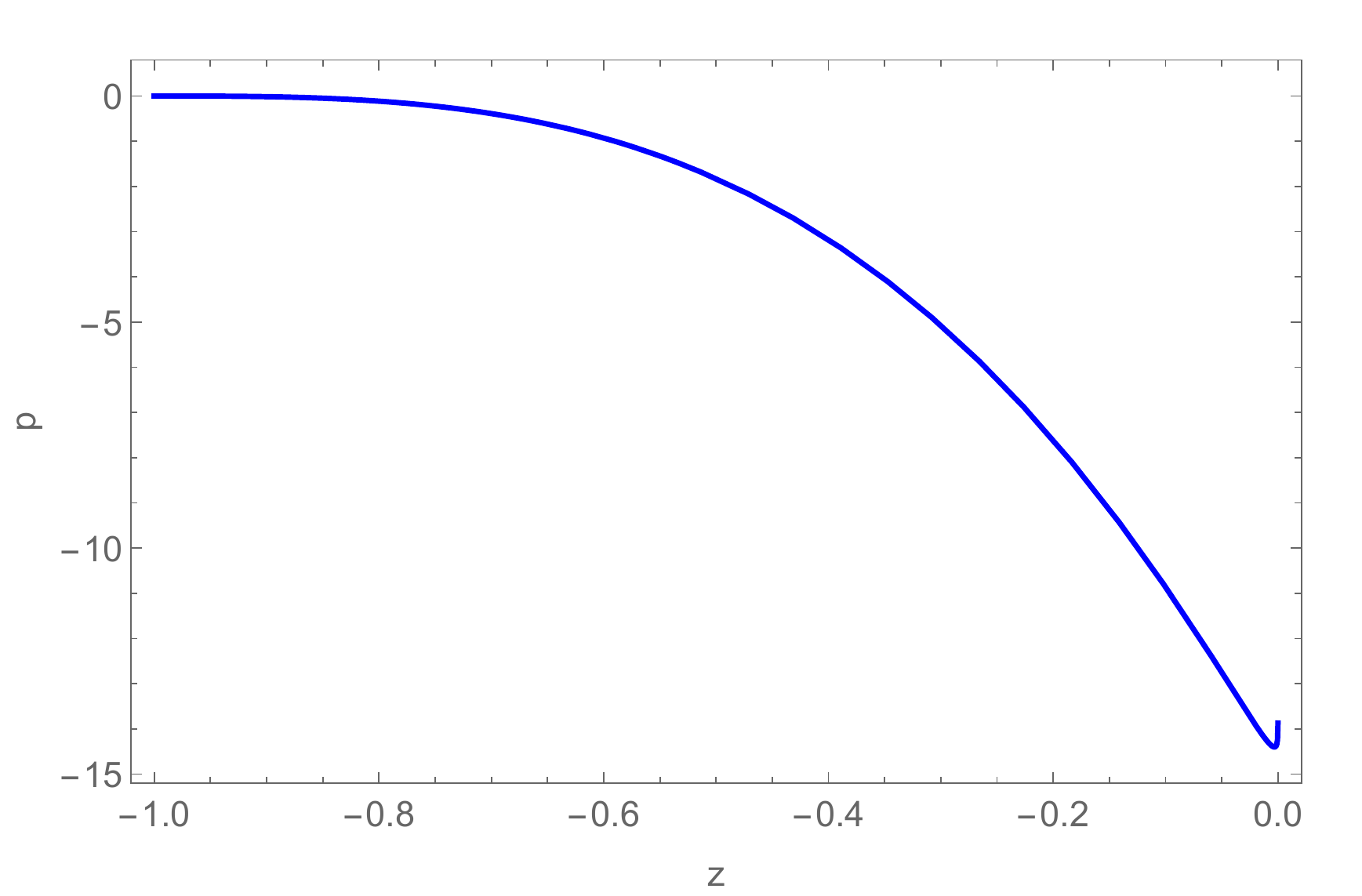}
\endminipage
\caption{Plot for the variation of the pressure vs. cosmic time (Left panel) and pressure vs. redshift (Right panel) for Model II with the $\rho_{c}=0.75$, $\lambda_1=-0.5$, $\lambda_2=-12.56$ and $m=1.01$.~~[Eq.\eqref{eq.24}]}
\minipage{0.50\textwidth}
\centering
\includegraphics[width=\textwidth]{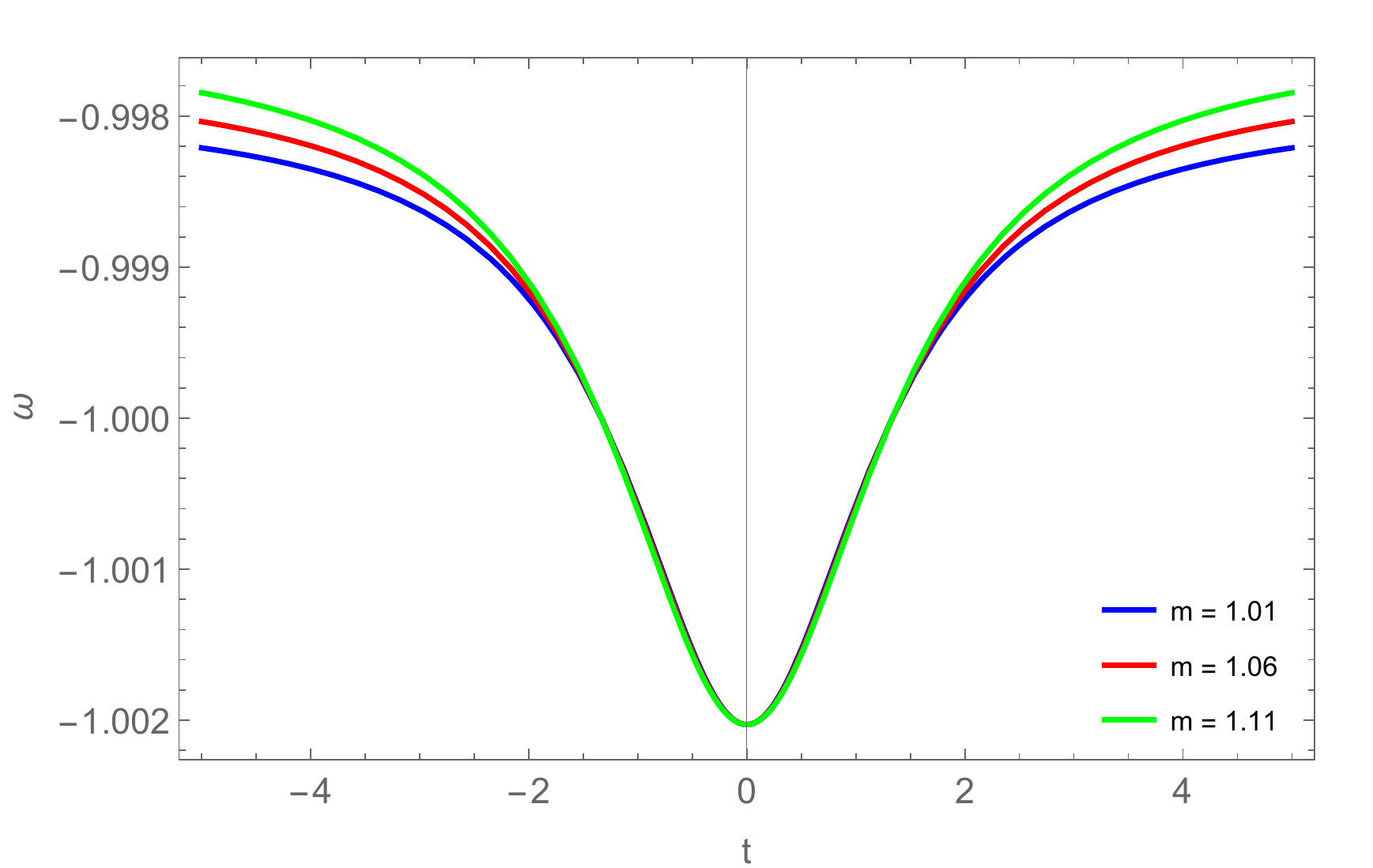}
\endminipage\hfill
\minipage{0.50\textwidth}
\includegraphics[width=\textwidth]{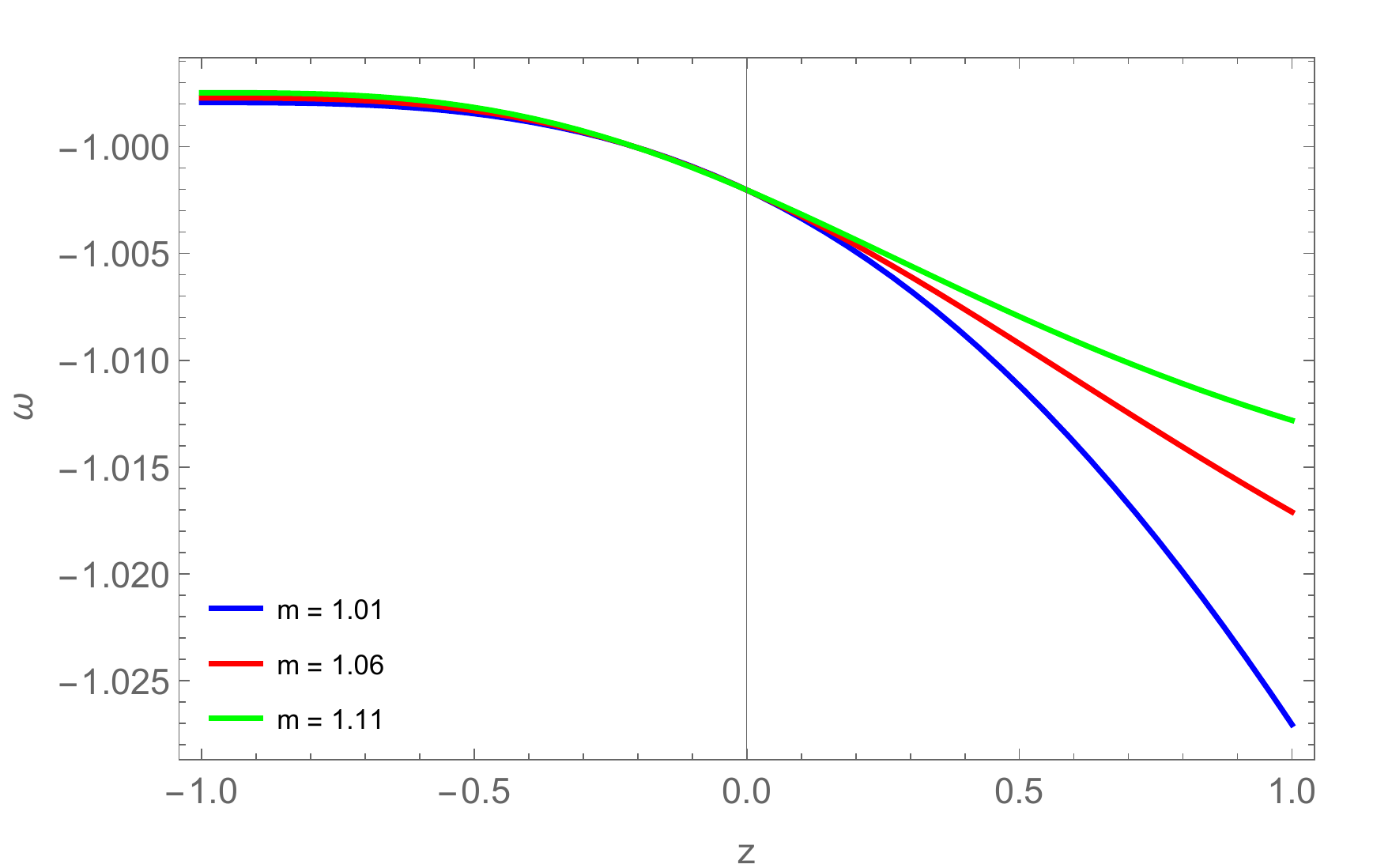} 
\endminipage
\caption{Plot for the variation of the EoS parameter vs. cosmic time (Left panel) and EoS parameter vs. redshift (Right panel) for Model II with the $\rho_{c}=0.75$, $\lambda_1=-0.5$, $\lambda_2=-12.56$ and $m=1.01, 1.06, 1.11$.~~[Eq.\eqref{eq.25}]}
\end{figure}

\section{Energy Conditions}

In Section-II, we have mentioned the  importance of the energy conditions in describing a model. Also, given the extension of the symmetric teleparallel gravity $f(Q,T)$ theory, we made a comprehensive discussion on what should be the additional conditions followed by a bouncing model. In fact, the prescribed condition to get the bouncing solution is the possible violation of null energy condition at bounce epoch. Here we wish to present the energy conditions for the two models discussed in the present work.

The energy conditions for Model-I may be expressed as,
\begin{equation} \label{eq.18}
\rho+p=\frac{2^m3^{m-1}m(1-2m)(1+z)^{2m\chi}\lambda_1\left[\chi-2\alpha(1+z)^{2\chi}\right]}{\tilde{\lambda}_2\chi^{(3m-1)}\left[\chi-\alpha(1+z)^{2\chi}\right]^{1-m}},
\end{equation}
\begin{equation} \label{eq.19}
\rho+3p=\frac{6^{m-1}(1-2m)(1+z)^{2m\chi}\lambda_1\left[\alpha (3\tilde{\lambda}_2 -\left(10\tilde{\lambda}_2-32\pi\right)m\chi)\chi(1+z)^{2\chi}+(\left(5\tilde{\lambda}_2-16\pi\right)m\chi-3\tilde{\lambda}_2)\right]}{{\chi^{3m}}\tilde{\lambda}_2\left(\tilde{\lambda}_2-4\pi\right)\left[\chi-\alpha(1+z)^{2\chi}\right]^{1-m}},
\end{equation}
\begin{equation} \label{eq.20}
\rho-p=\frac{6^{m-1}(1-2m)(1+z)^{2m\chi}\lambda_{1}\left[\alpha(2m\chi-3)\chi (1+z)^{2\chi}-(m\chi-3)\right]}{\chi^{3m}(\tilde{\lambda}_2-4 \pi)\left[\chi-\alpha(1+z)^{2\chi}\right]^{1-m}}.
\end{equation}\\

The energy conditions for Model- II are expressed as,
\begin{equation} \label{eq.26}
\rho+p=\frac{m2^{m-1}(1-2m)\rho_{c}^{m}(1+z)^{3m}(-z)^{m}\lambda_1\left[2z^{3}+6z^{2}+6z+1\right](z^{2}+3z+3)^{m}}{\tilde{\lambda}_2(z^{3}+3z^{2}+3z),}
\end{equation}
\begin{equation} \label{eq.27}
\rho+3p=\frac{2^{m-2}(1-2m)\rho_{c}^{m}\lambda_1\left[m(5\tilde{\lambda}_2-16\pi)(2z^{3}+6z^{2}+6z+1)-2\tilde{\lambda}_2(z^{3}+3z^{2}+3z)\right]}{\tilde{\lambda}_2\left(\tilde{\lambda}_2-4\pi\right)(z^{3}+3z^{2}+3z)[(z+1)^{3}-(1+z)^{6}]^{-m}},
\end{equation}
\begin{equation} \label{eq.28}
\rho-p=-\frac{2^{m-2}(1-2m)\rho_{c}^{m}\lambda_1\left[2(m-1)(z^{3}+3z^{2}+3z)+m\right]}{\left(\tilde{\lambda}_2-4\pi\right)(z^{3}+3z^{2}+3z)[(z+1)^{3}-(1+z)^{6}]^{-m}}.
\end{equation}

We show the energy condition violation in FIG.9 for both the models. In the left panel we show the evolution of the energy conditions in Model I and in the right panel, the same for the Model II is presented. Also, in the figures, we show that dominant energy condition (black or blue line) is satisfied and the strong energy condition (green line) is violated. However, the null energy condition (red line) is marginally violated near the bounce.  Moreover, the obvious phenomenon is the violation of the strong energy condition when the null energy condition gets violated. However, from the graphical representation, it has been observed that the strong energy condition gets violated throughout the evolution. It is worthy to mention that, the violation of the null energy condition in the early Universe is linked to the non singular bouncing solution \cite{Cai07}. In the present work, the same has been observed that supports the bouncing solution of the Universe.

\begin{figure}[tbph]
\minipage{0.50\textwidth}
\centering
\includegraphics[width=\textwidth]{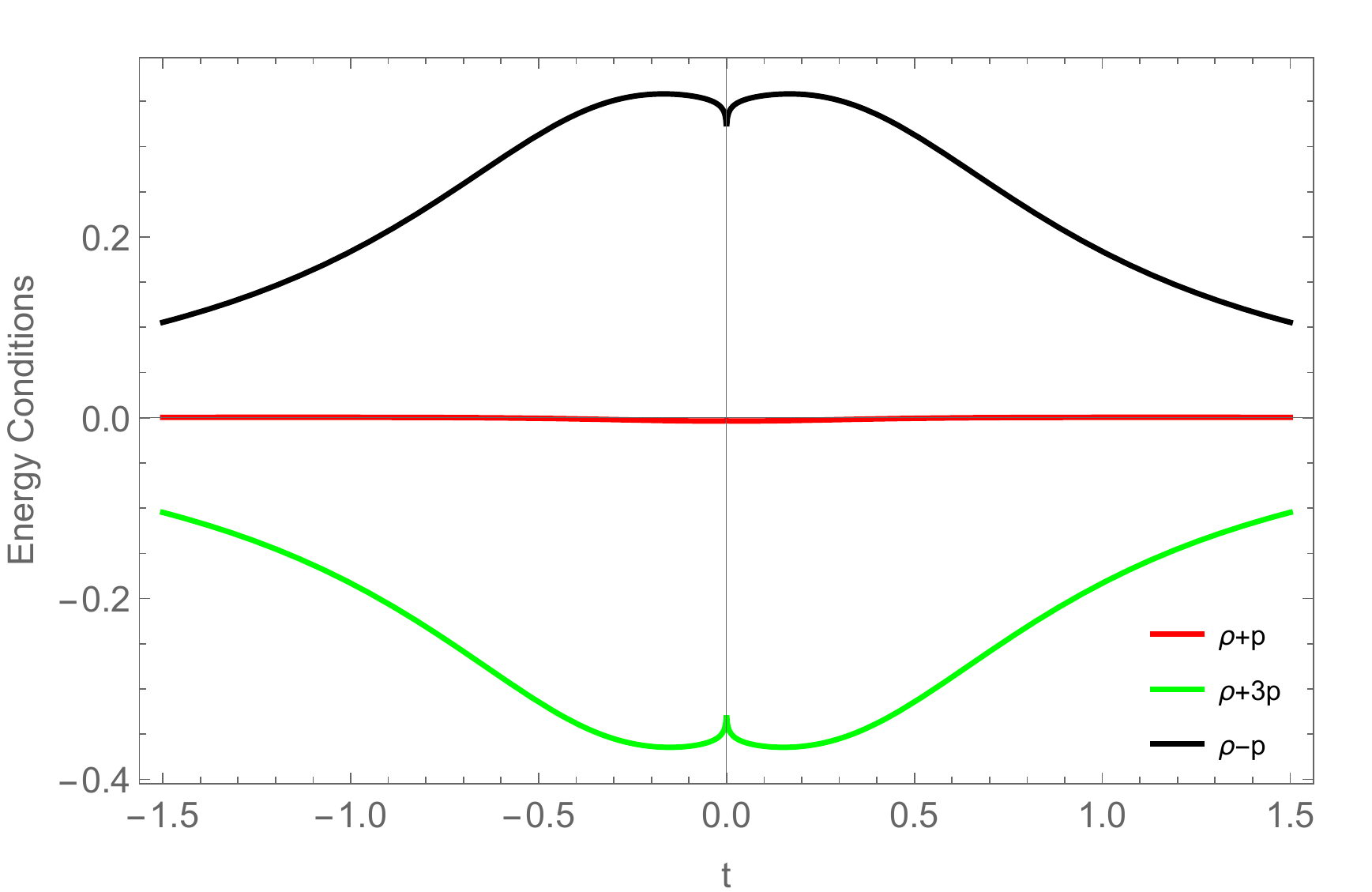}
\endminipage\hfill
\minipage{0.50\textwidth}
\includegraphics[width=\textwidth]{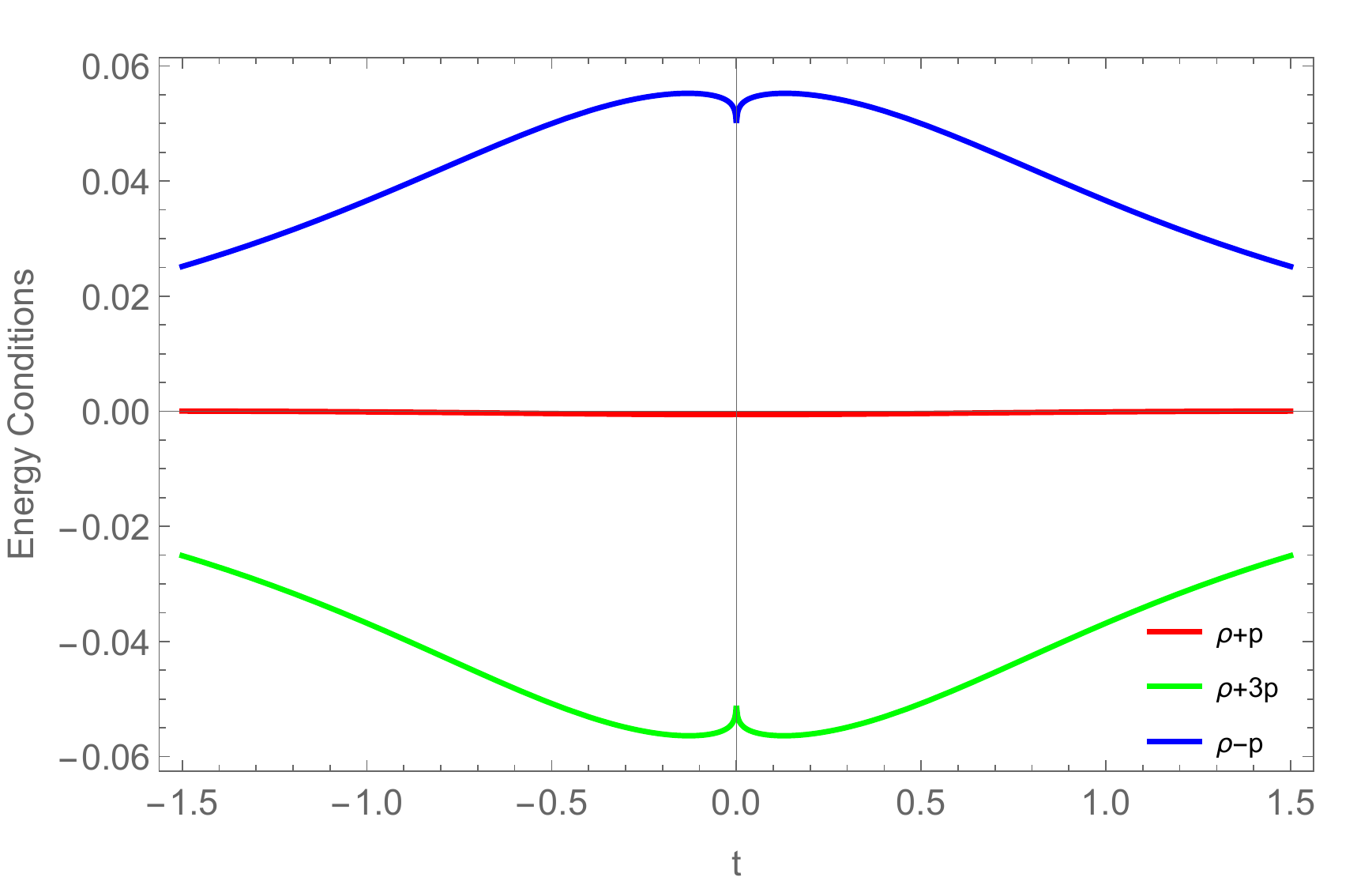} 
\endminipage
\caption{Energy conditions. (a)(Left panel)  plot of the energy conditions in Model I. The model parameters are chosen as $\alpha=0.43$ and $\chi=1.001$ (b) (Right panel) same for Model II with the model parameter $\rho_{c}=0.75$. For both the models , we chose the $f(Q,T)$ parameters as $\lambda_1=-0.5$, $\lambda_2=-12.5$ and $m=1.01$.}
\end{figure}



\section{Validation through Cosmographic test}

According to the cosmological principle, the scale factor seems to be the only degree of freedom governing the Universe. We may expand the scale factor $a(t)$ in a Taylor series around $a_0$ as \cite{Aviles2012, Gruber2014, SKT2019},
\begin{eqnarray}
a(t)=a_0+\sum_{n=0}^{\infty} \frac{1}{n!}\frac{d^na}{da^n}\mid_{(t=t_0)} (t-t_0)^n,\label{eq:32}
\end{eqnarray}
where $t_0$ is the present cosmic time and $n=1,2,3, \cdots$ is an integer. The coefficients of expansion are referred to as the cosmographic coefficients. In fact, the cosmographic coefficients contain different order of derivatives of the scale factor and therefore may provide a better geometrical description of the model. At any time $t$, we may define some geometrical parameters as,
\begin{eqnarray}
H &=& \frac{\dot{a}}{a}\\
q &=& -\frac{a\ddot{a}}{\dot{a}^2}\\
j &=& \frac{\dddot{a}}{aH^3}\\
s &=& \frac{a^{(4)}}{aH^4}\\
l &=& \frac{a^{(5)}}{aH^5},
\end{eqnarray}
where $a^{(4)}$ and $a^{(5)}$ are respectively the fourth order and fifth order derivatives of the scale factor. $H$ and $q$ are the Hubble parameter and the deceleration parameter. $j$ and $s$ are called the jerk parameter and snap parameter respectively and form the state finder pairs. $l$ is known as the lerk parameter. The state finder pair are expressed in terms of the deceleration parameter as,
\begin{eqnarray}
j &=& q+2q^2-\frac{\dot{q}}{H}\\
s &=& \frac{j-1}{3(q-\frac{1}{2})}.
\end{eqnarray}
For $\Lambda$CDM model, the state finder pair $\{j,s\}$ assume the values $\{1,0\}$. It should be mentioned here that, while the state finder pair have been constrained to some extent, the value of the lerk parameter is to be fixed from observations. Owing to the large uncertainty in the observations of supernova at very high redshift, it becomes difficult to pin down the values of these cosmographic parameters. We have already discussed about the Hubble parameter in the respective bouncing models. Here, we will discuss the remaining parameters. The sign of the deceleration parameter $q$  determines whether the Universe is accelerating or decelerating. In other words, a positive deceleration parameter indicates that standard gravity predominates over the other species, whereas a negative sign provides a repulsive effect that overcomes the standard attraction due to gravity. The sign of $j$ defines the change in the dynamics of the Universe, with a positive value signifying the occurrence of a transition phase when the Universe adjusts its expansion. Furthermore, the value of $s$ is required to distinguish between a dark energy term that is evolving and a cosmological constant behaviour. The functional dependence of dark energy on the redshift $z$ is affected by departures from the predicted value of $s$, which is evaluated in the concordance model, demonstrating that it evolves as the Universe expands. The terms $s$ and $l$ both impact higher orders of the Taylor expansion at greater redshifts \cite{Dunsby16}. One should note that, these cosmographic parameters have nothing to do with the $f(Q,T)$ gravity theory or the choice of the parameters $m, \lambda_1$ and $\lambda_2$. The cosmographic parameters simply depend on the constant parameters appearing in the scale factor. Since we have considered two different ansatz for the scale factor representing a matter bounce scenario, in the following we present the expressions for the cosmographic parameters of the two models separately. Also, we wish to express these parameters in terms of the redshift.

The cosmographic parameters for the Model I are obtained as,
\begin{eqnarray} \label{eq.21}
q(z)&=&-1-\frac{\alpha \chi (1+z)^{2\chi}}{\chi-\alpha(1+z)^{2\chi}}+\chi, \nonumber \\
j(z)&=&-1+\frac{ 3\alpha \chi (1-2\chi)(1+z)^{2\chi}}{\chi-\alpha(1+z)^{2\chi}}+\chi, \nonumber \\
s(z)&=&\left(\frac{3\alpha^{2}\chi^{2}(1+z)^{4\chi}}{(\chi-\alpha(1+z)^{2\chi})^{2}}+\frac{6\alpha \chi(1-3\chi)(1+z)^{2\chi}}{\chi-\alpha(1+z)^{2\chi}}+(\chi-1)(3\chi-1)\right)(2\chi-1), \nonumber \\
l(z)&=&\left(\frac{15\alpha^{2}\chi^{2}(1+z)^{4\chi}}{(\chi-\alpha(1+z)^{2\chi})^{2}}+\frac{10\alpha \chi(1-3\chi)(1+z)^{2\chi}}{\chi-\alpha(1+z)^{2\chi}}+(\chi-1)(3\chi-1)\right)(8\chi^{2}-6\chi+1).
\end{eqnarray}\\

The cosmographic parameters for the Model II are obtained as 

\begin{eqnarray} \label{eq.29}
q(z)&=&\frac{1}{2}-\frac{3}{2}\frac{(1+z)^{3}}{(1-(1+z)^{3})}, \nonumber \\
j(z)&=&1-\frac{9(1+z)^{3}}{(1-(1+z)^{3})}, \nonumber \\
s(z)&=&-\frac{7}{2}-\frac{27}{2}\frac{(1+z)^{6}}{(1-(1+z)^{3})^{2}}+\frac{21}{4}\frac{(1+z)^{3}}{(1-(1+z)^{3})}, \nonumber \\
l(z)&=&\frac{35}{2}+\frac{675}{2}\frac{(1+z)^{6}}{(1-(1+z)^{3})^{2}}-525\frac{(1+z)^{3}}{(1-(1+z)^{3})}.
\end{eqnarray}
\begin{figure}
    \centering
    \subfigure[]{\includegraphics[width=4.2cm,height=3.5cm]{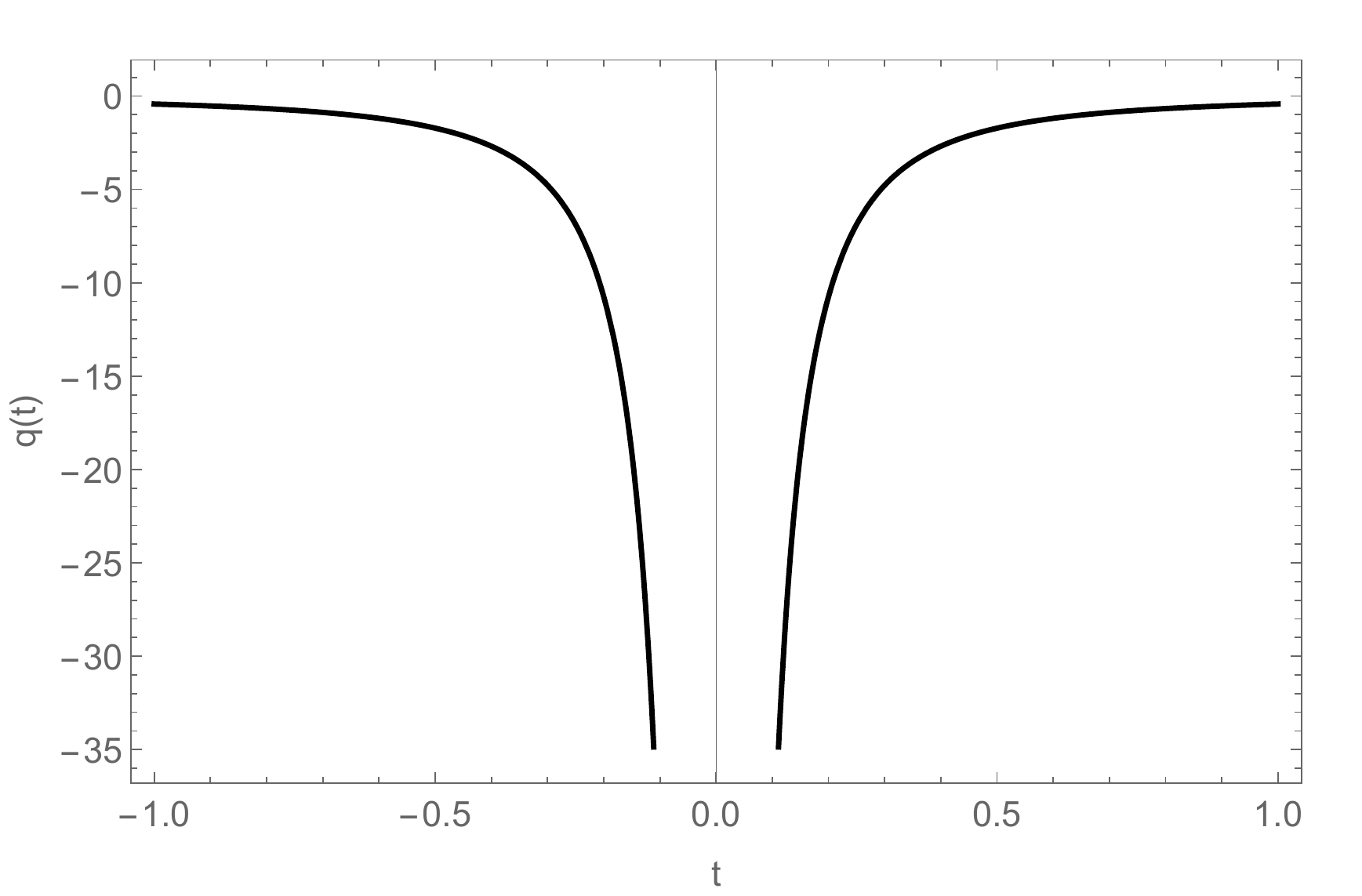}} 
    \subfigure[]{\includegraphics[width=4.2cm,height=3.5cm]{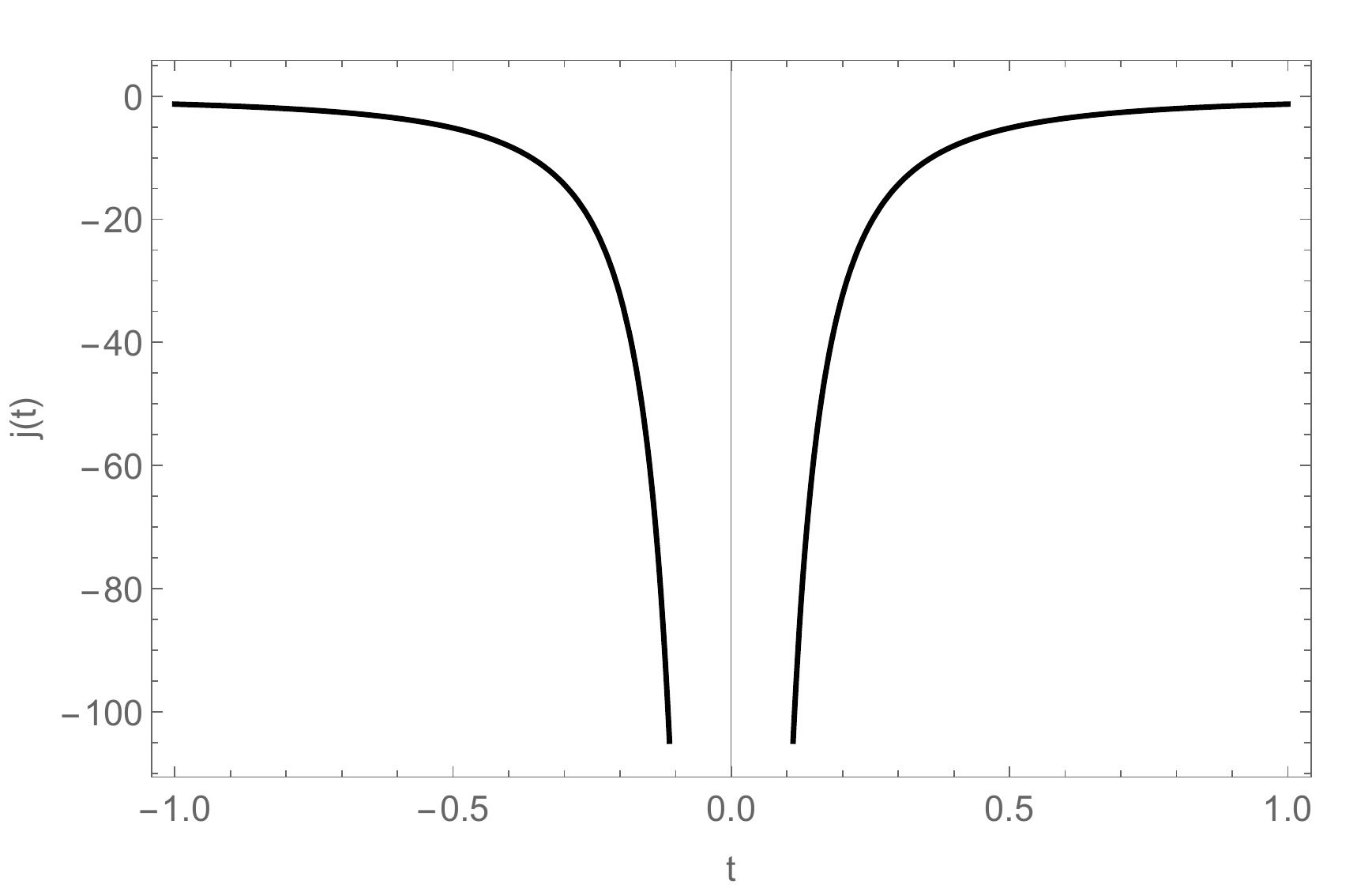}} 
    \subfigure[]{\includegraphics[width=4.2cm,height=3.5cm]{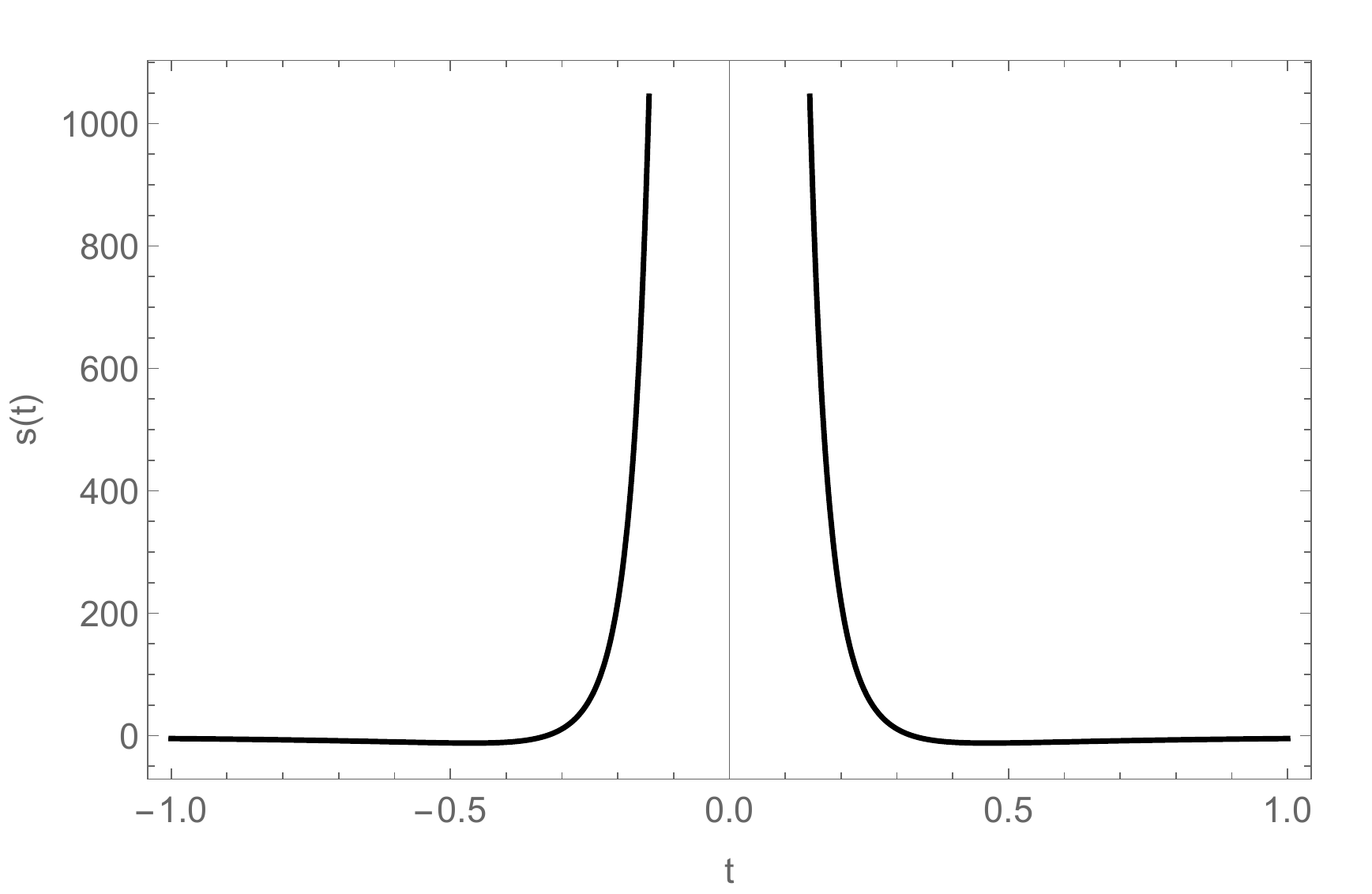}}
    \subfigure[]{\includegraphics[width=4.2cm,height=3.5cm]{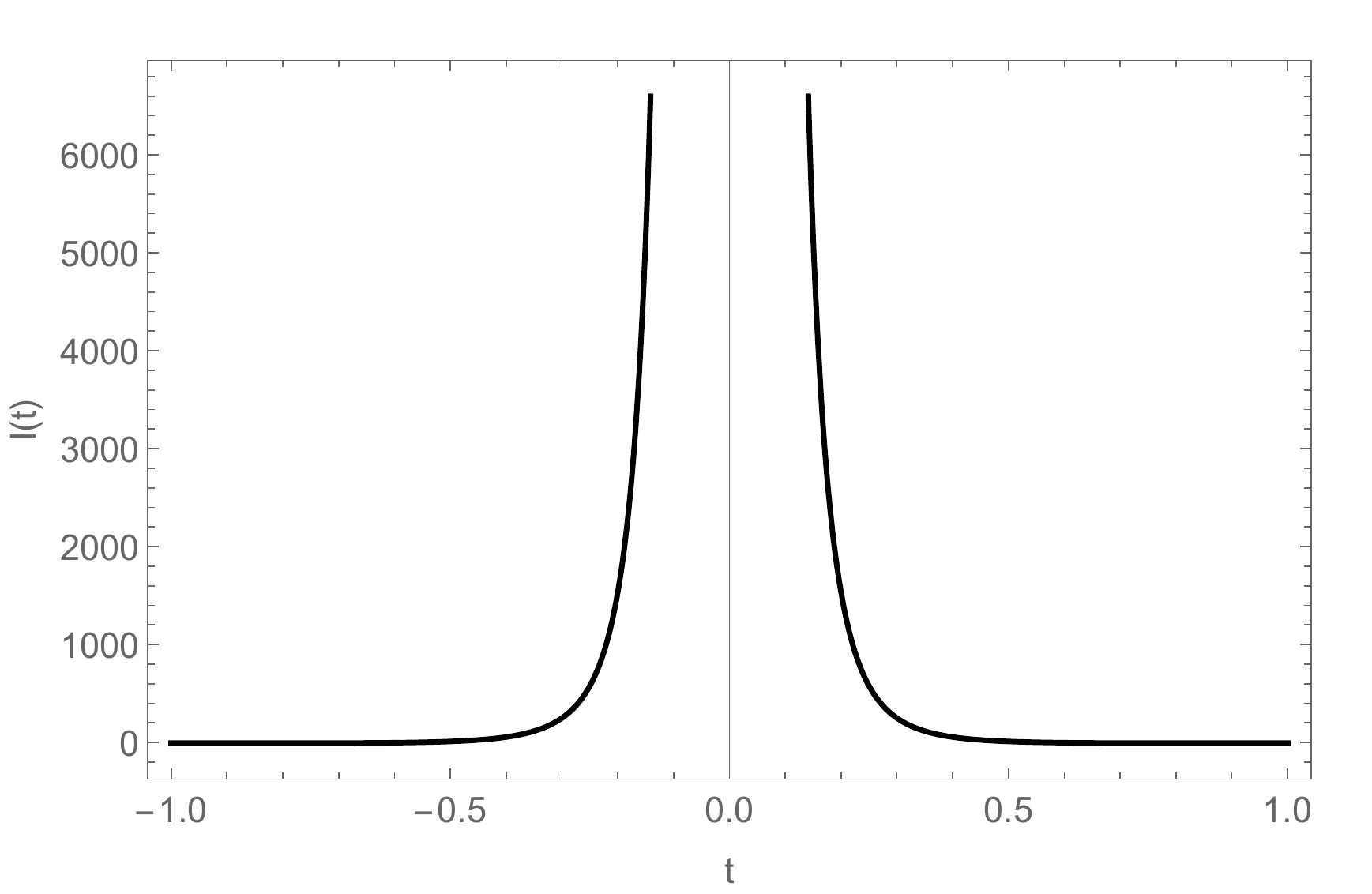}}
    \caption{Plot for the variation of (a) the deceleration parameter, (b) jerk parameter, (c) snap parameter  and (d) the lerk parameter as functions of  cosmic time for Model I with $\alpha=0.43$ and $\chi=1.001$.~~[Eq.\eqref{eq.21}]}
\end{figure}

\begin{figure}
    \centering
    \subfigure[]{\includegraphics[width=4.2cm,height=3.5cm]{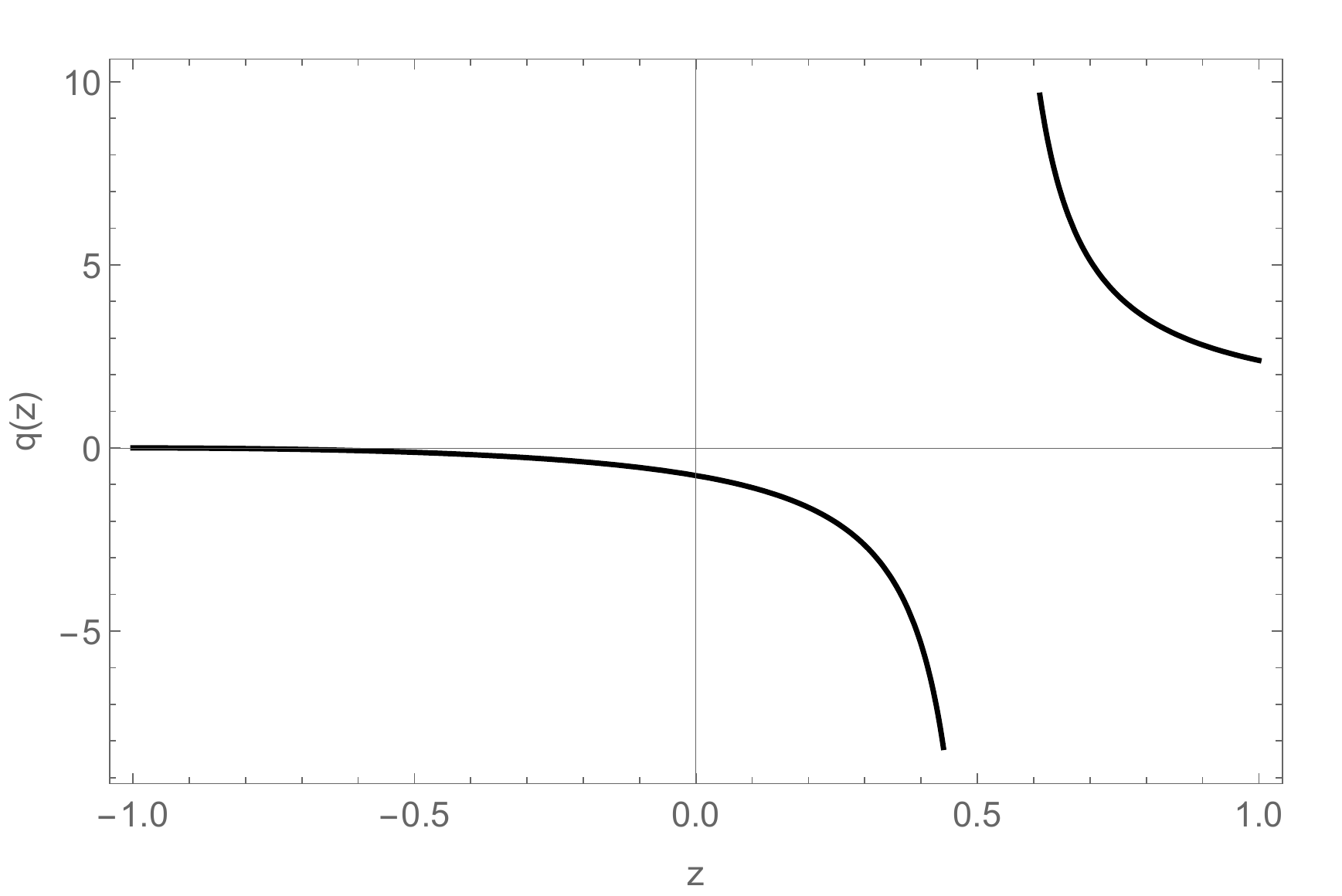}} 
    \subfigure[]{\includegraphics[width=4.2cm,height=3.5cm]{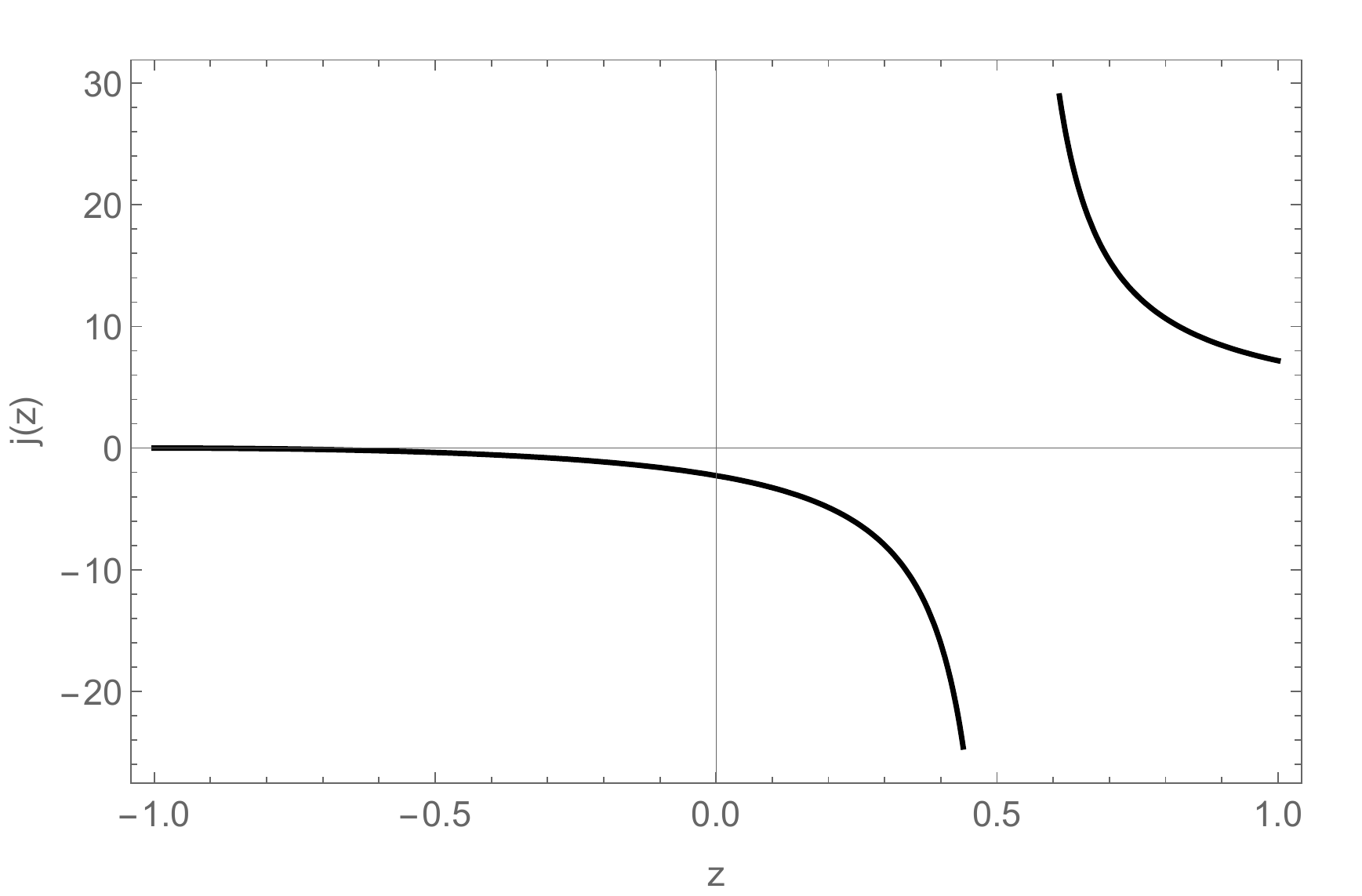}} 
    \subfigure[]{\includegraphics[width=4.2cm,height=3.5cm]{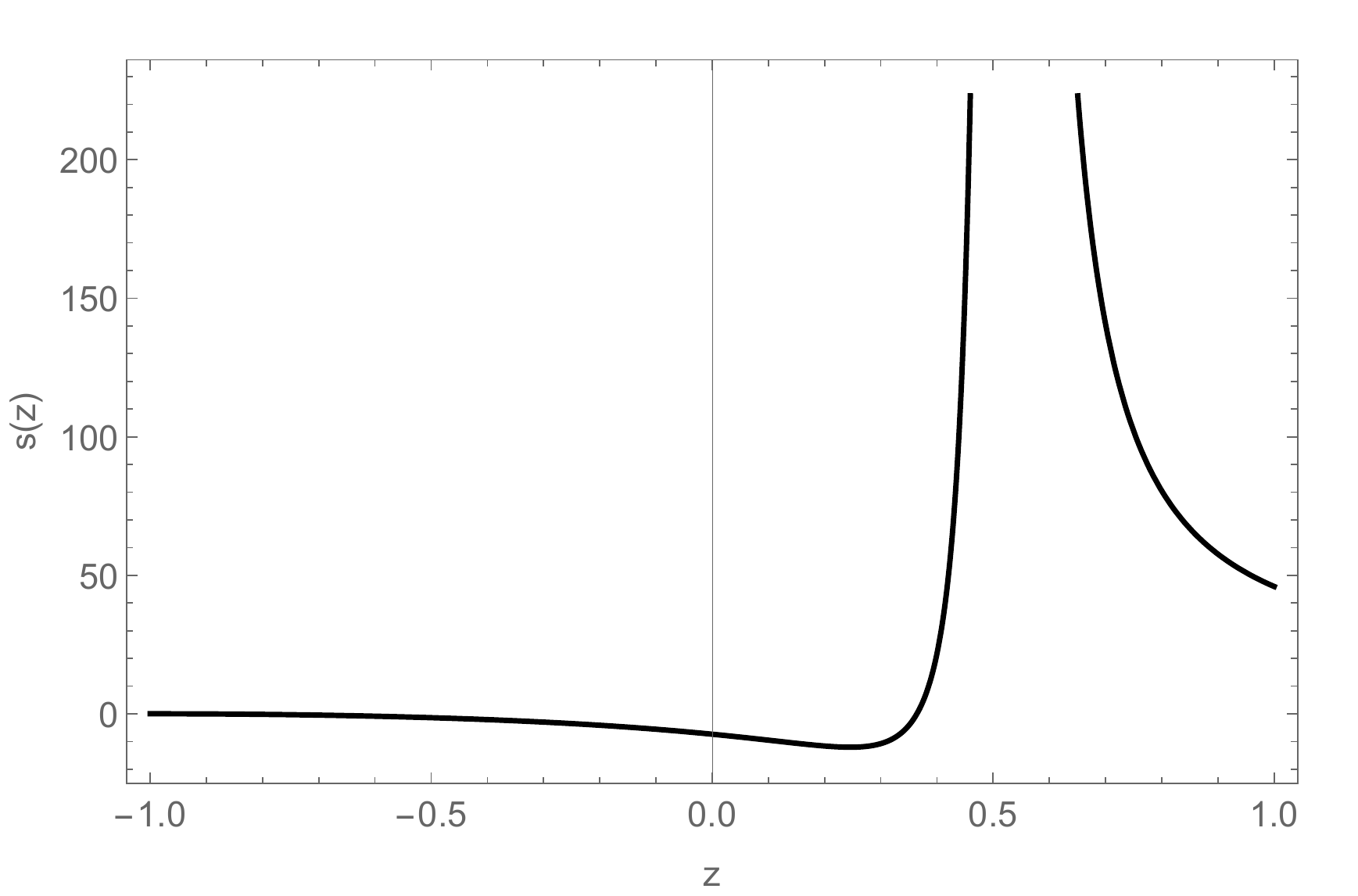}}
    \subfigure[]{\includegraphics[width=4.2cm,height=3.5cm]{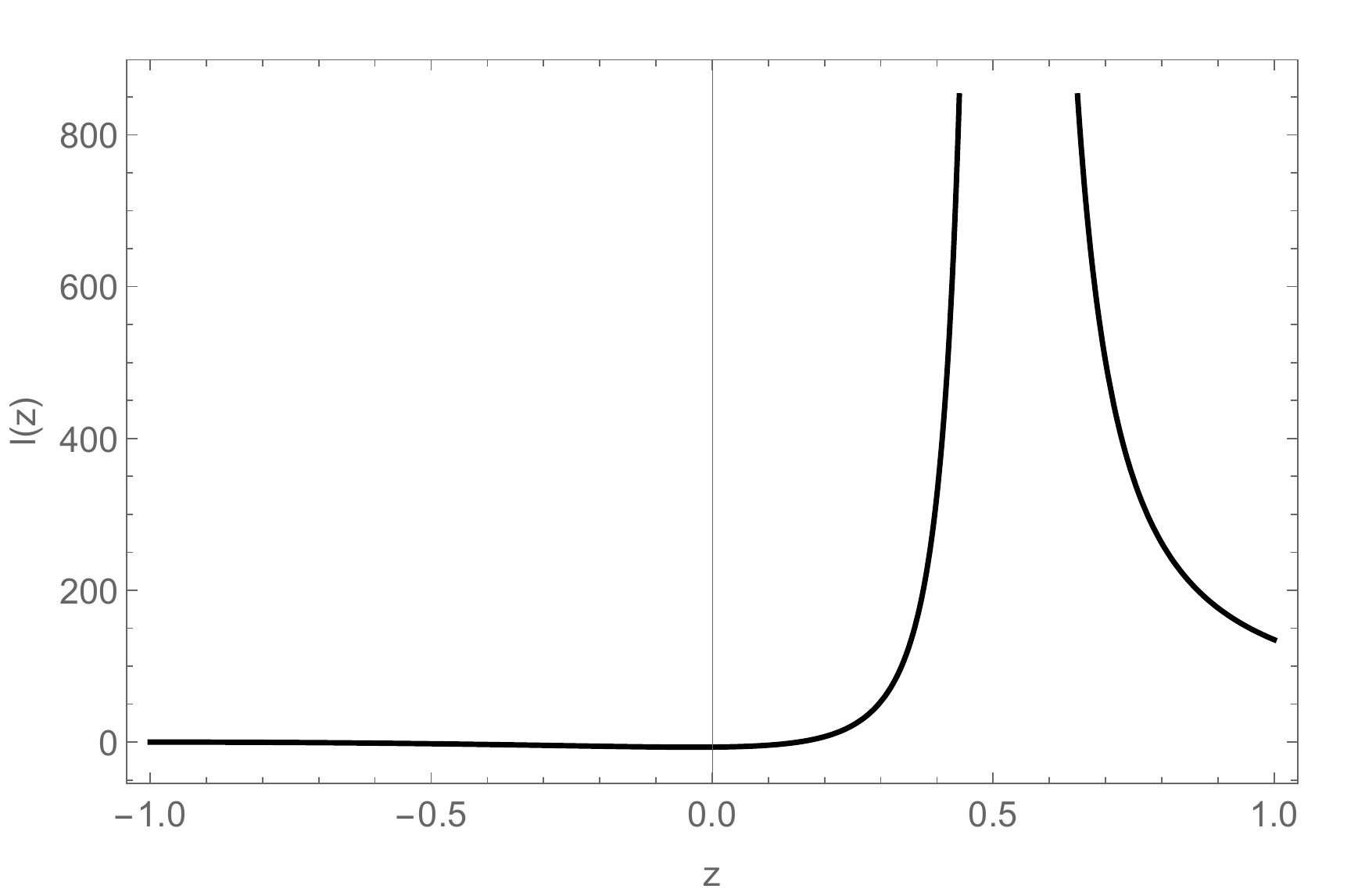}}
    \caption{Plot for the variation of (a) the deceleration parameter, (b) jerk parameter, (c) snap parameter and (d) the lerk parameter as functions of redshift for Model I with $\alpha=0.43$ and $\chi=1.001$.~~[Eq.\eqref{eq.21}]}
\end{figure}


For both the models, the cosmographic parameters evolve with cosmic time and thereby signify their importance in the context of an evolving dark energy form in the Universe. Of course, in the present extended symmetric teleparallel gravity theory, the dark energy is chosen to be strictly geometric in nature. One interesting aspect in these expressions is that, while the cosmographic parameters in Model I depend on the scale factor parameters $\chi$ and $\alpha$, in the second model, they are independent of the scale factor parameter $\rho_c$. FIG. 10 and FIG. 11 provide the graphical representations of the deceleration, jerk, snap, and lerk parameters with cosmic time and redshift respectively. In both the negative and positive time zones, the deceleration parameter is negative, with a singularity at the bounce point and convergence to $-0.1$ as it moves away from the bounce epoch. The jerk and snap parameters, on the other hand, have a unique negative behaviour. The jerk parameter has a negative value throughout the evolution, whereas the snap parameter has opposite values and demonstrates a positive behaviour away from the bounce point. The lerk parameter has a singularity in its positive regime; it goes through a brief period of negative values before reverting to zero as it moves away from the bounce point. For different scales, the deceleration and jerk parameters with the redshift parameter have the same behaviour; at $z=0.52$, it has transient behaviour passing through the singularity and reaching zero from a negative profile. The snap and lerk parameters, on the other hand, have a singularity near the bounce. The snap parameter begins with a negative value and decreases until it reaches the singularity, whereas the lerk parameter begins with a positive value and increases until it reaches the singularity. The order is shifted once more, and both approach to zero as $z$ tends to $-1$. In FIG. 12 and FIG. 13, the graphical representations of the deceleration parameter $q$, jerk parameter $j$, the snap parameter $s$ and lerk parameter $l$ are shown as functions of cosmic time and redshift respectively for Model II. The behaviour of these parameters are almost the same as those discussed in Model I.   


\begin{figure}
    \centering
    \subfigure[]{\includegraphics[width=4.2cm,height=3.5cm]{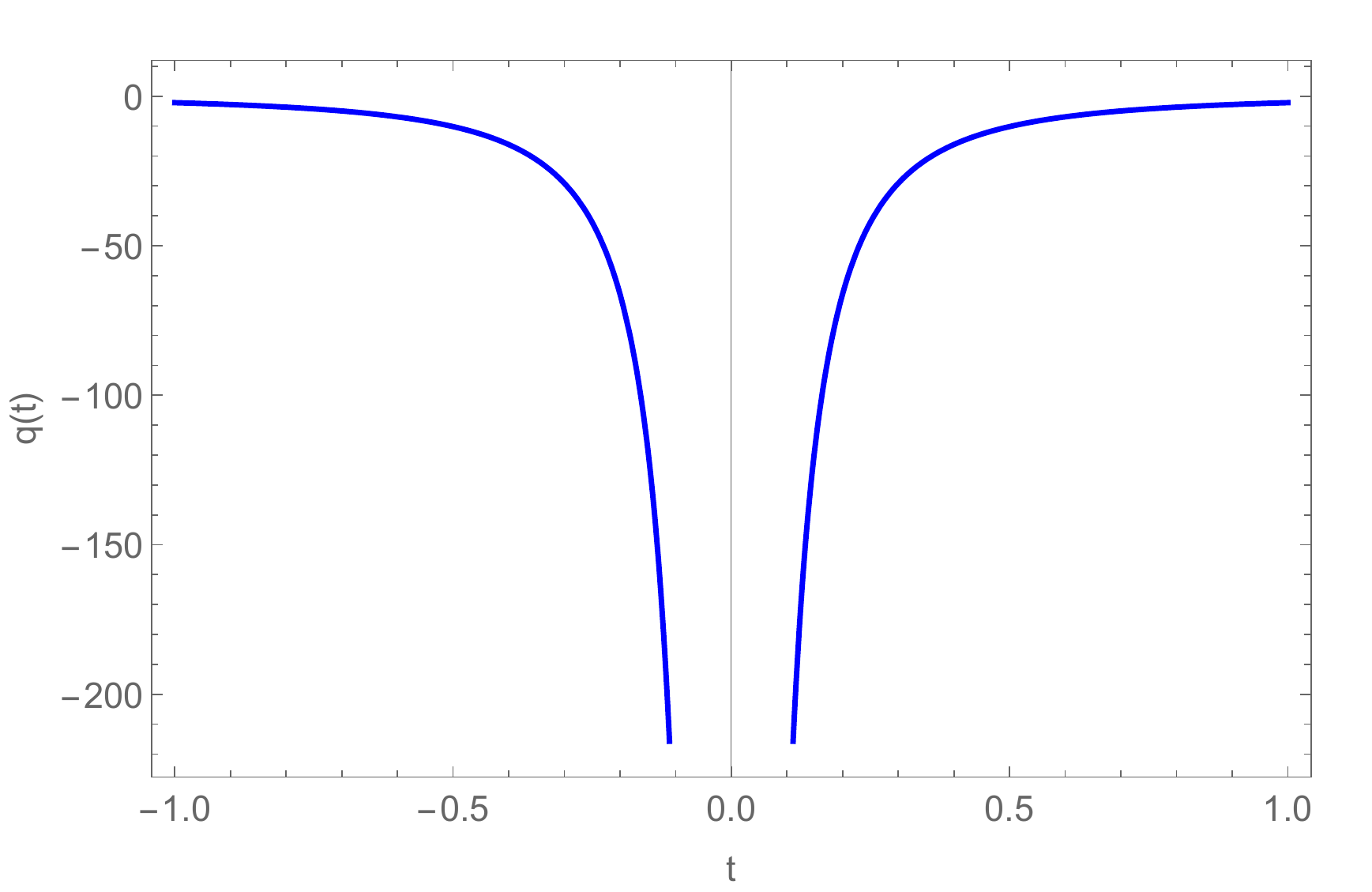}} 
    \subfigure[]{\includegraphics[width=4.2cm,height=3.5cm]{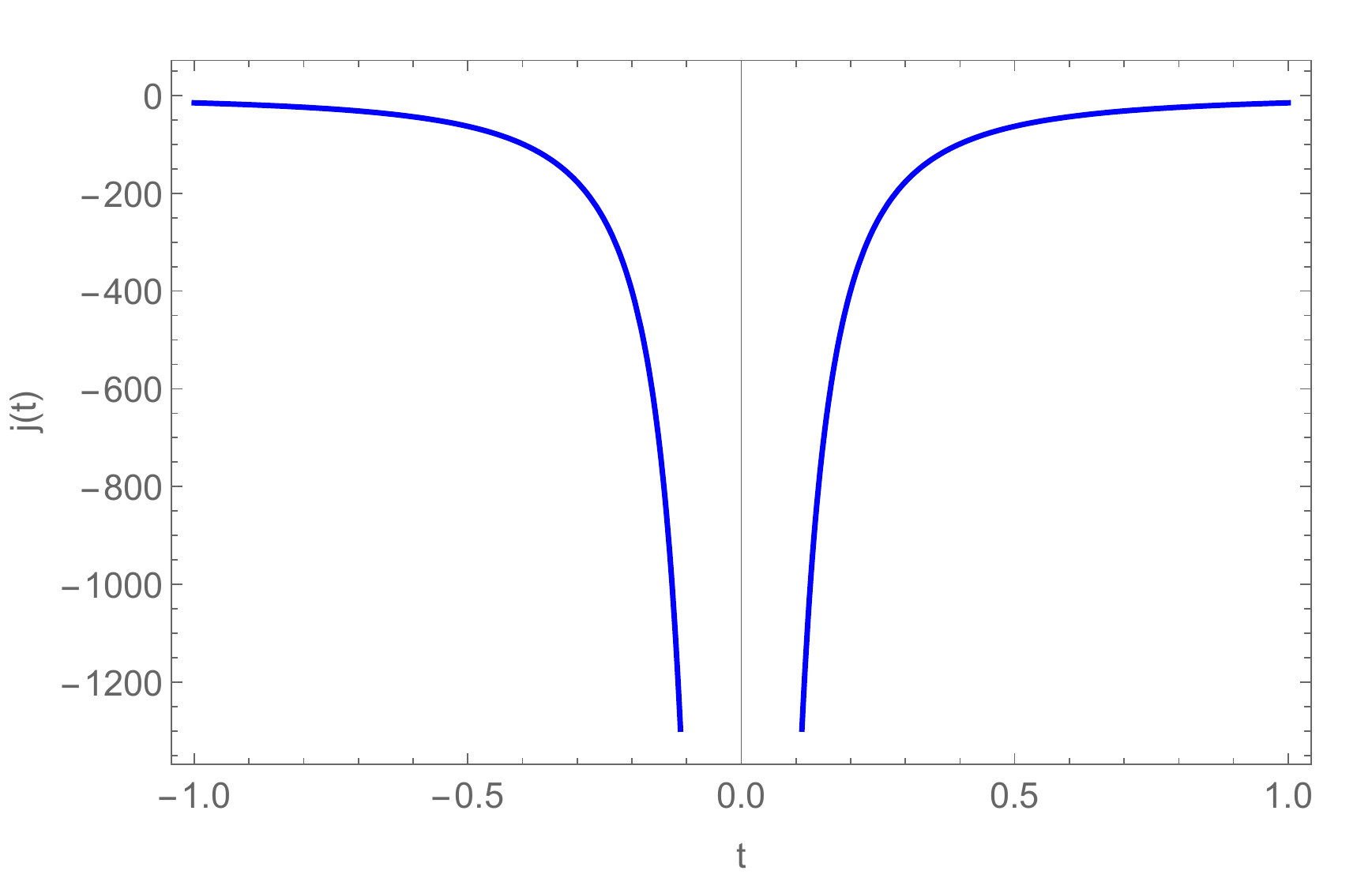}} 
    \subfigure[]{\includegraphics[width=4.2cm,height=3.5cm]{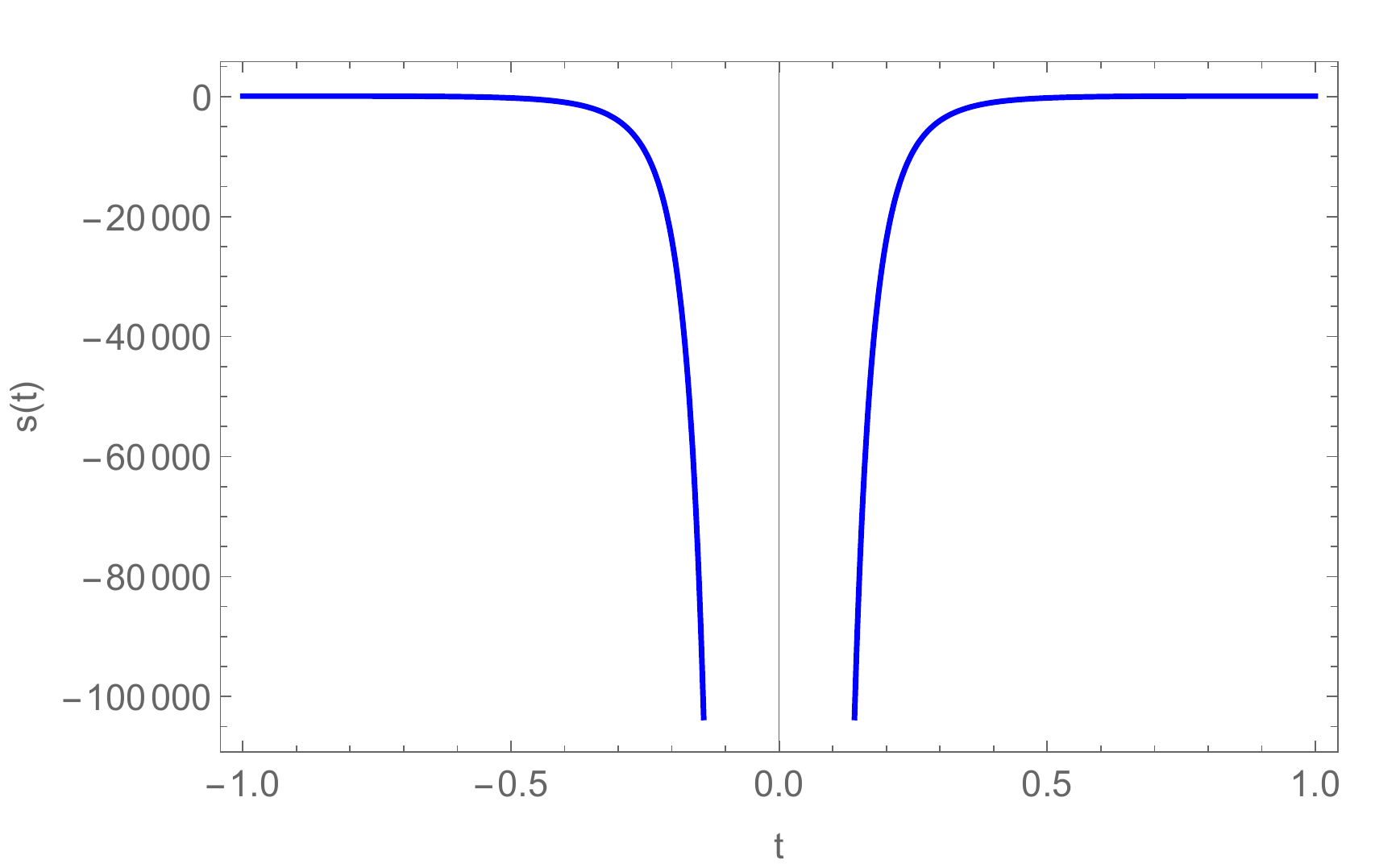}}
    \subfigure[]{\includegraphics[width=4.2cm,height=3.5cm]{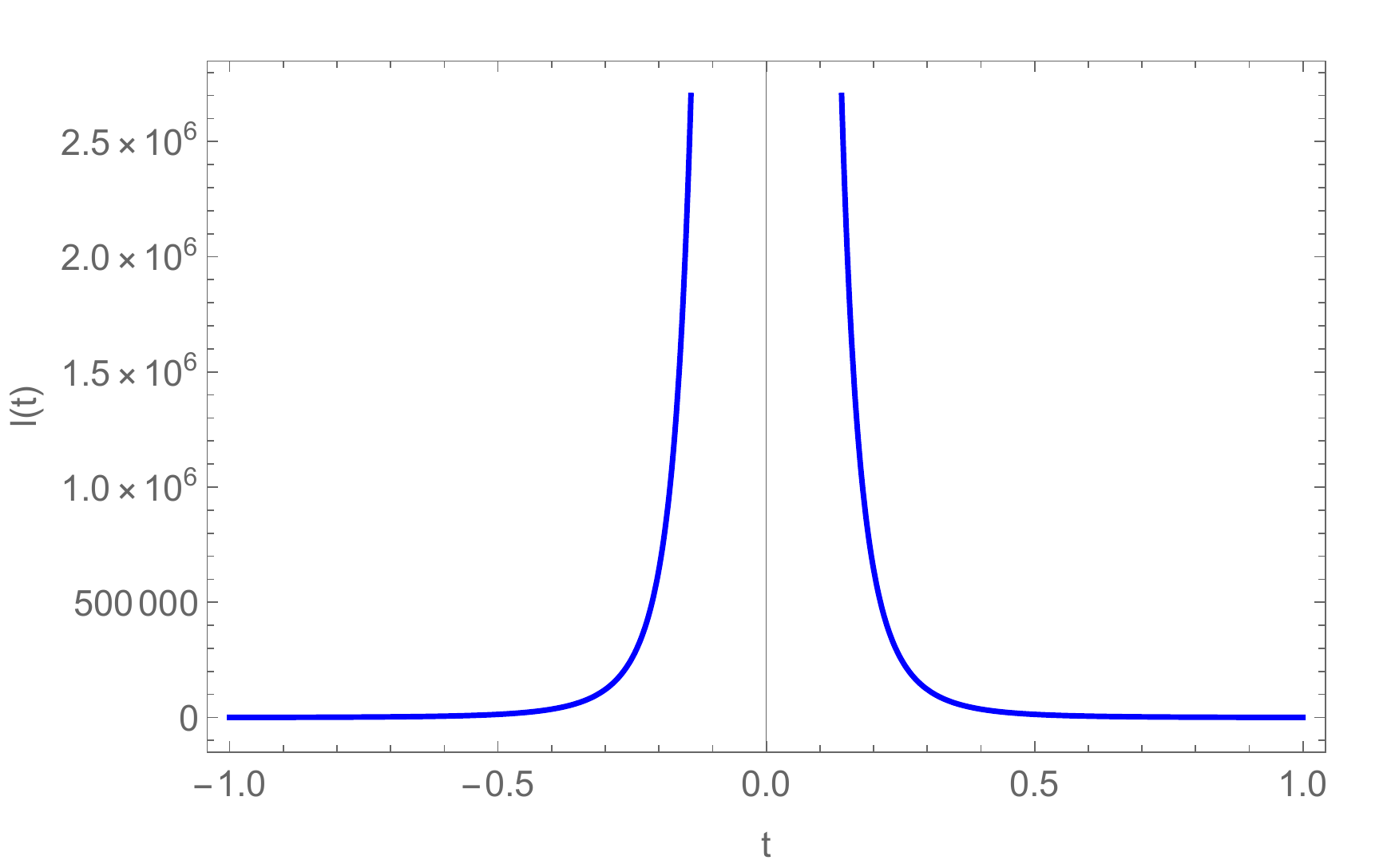}}
    \caption{Plot for the variation of (a) the deceleration parameter, (b) jerk parameter, (c) snap parameter  and (d) the lerk parameter as functions of  cosmic time for Model II with $\rho_{c}=0.75$.~~[Eq.\eqref{eq.29}]}
\end{figure}
\begin{figure}
    \centering
    \subfigure[]{\includegraphics[width=4.2cm,height=3.5cm]{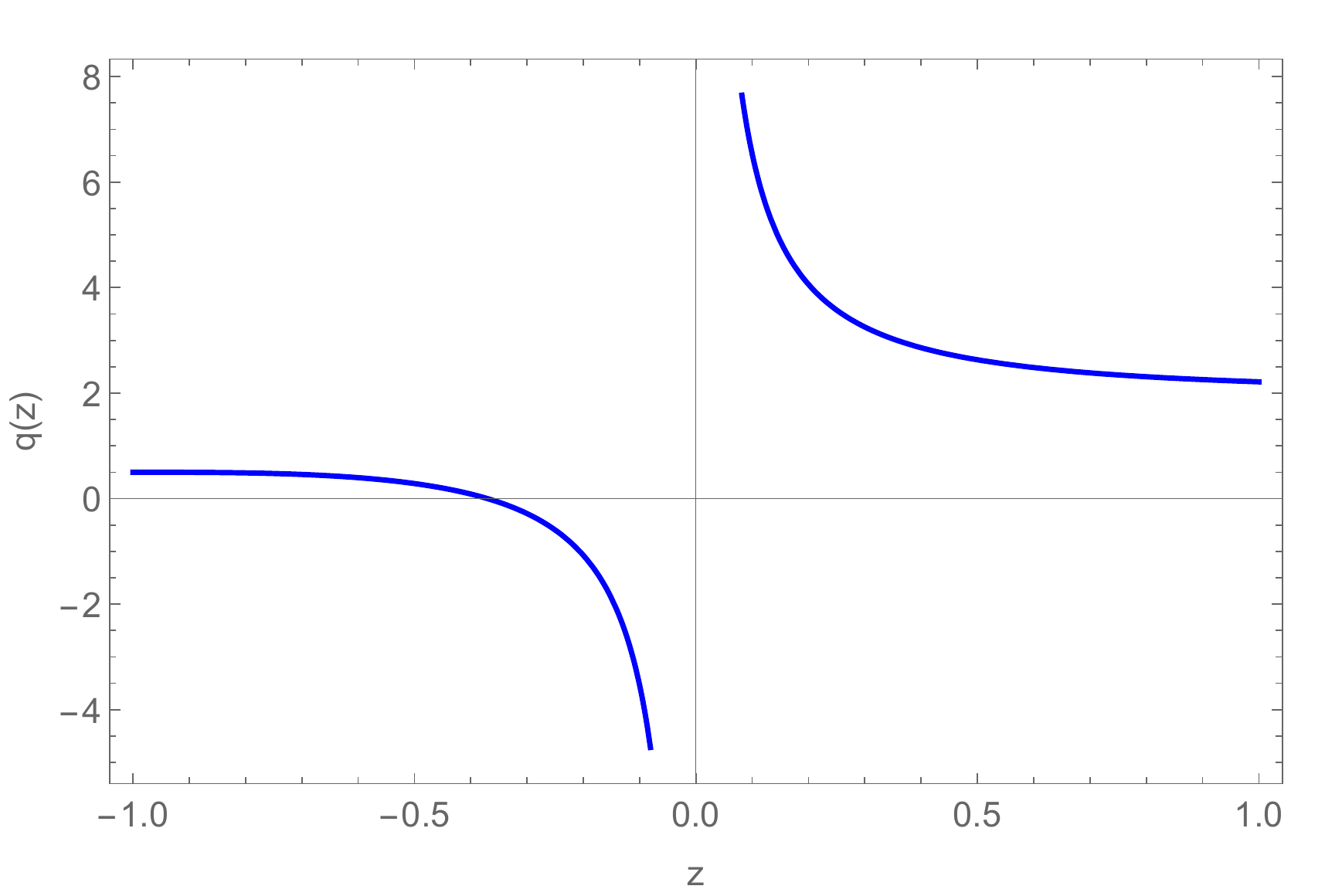}} 
    \subfigure[]{\includegraphics[width=4.2cm,height=3.5cm]{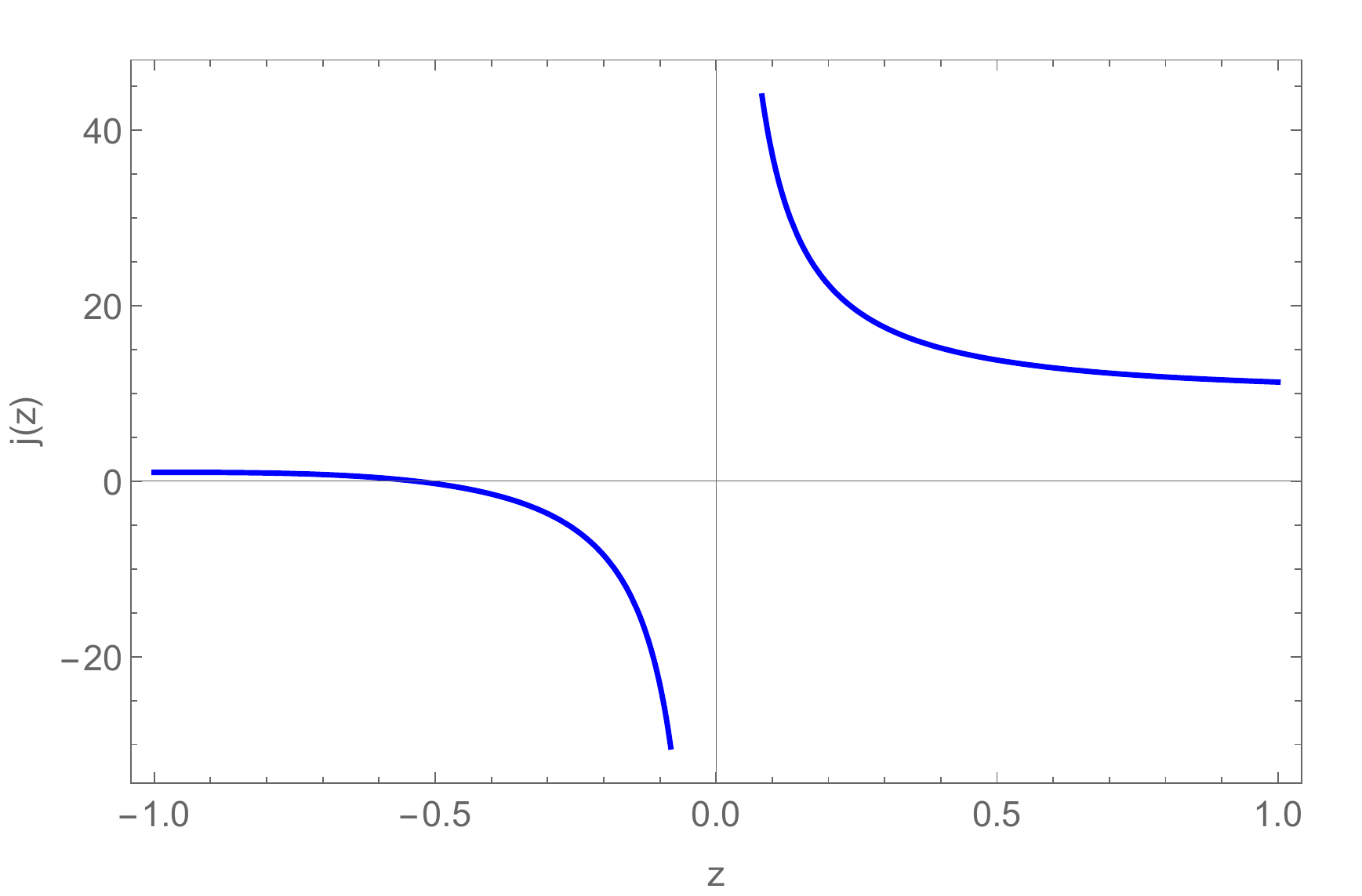}} 
    \subfigure[]{\includegraphics[width=4.2cm,height=3.5cm]{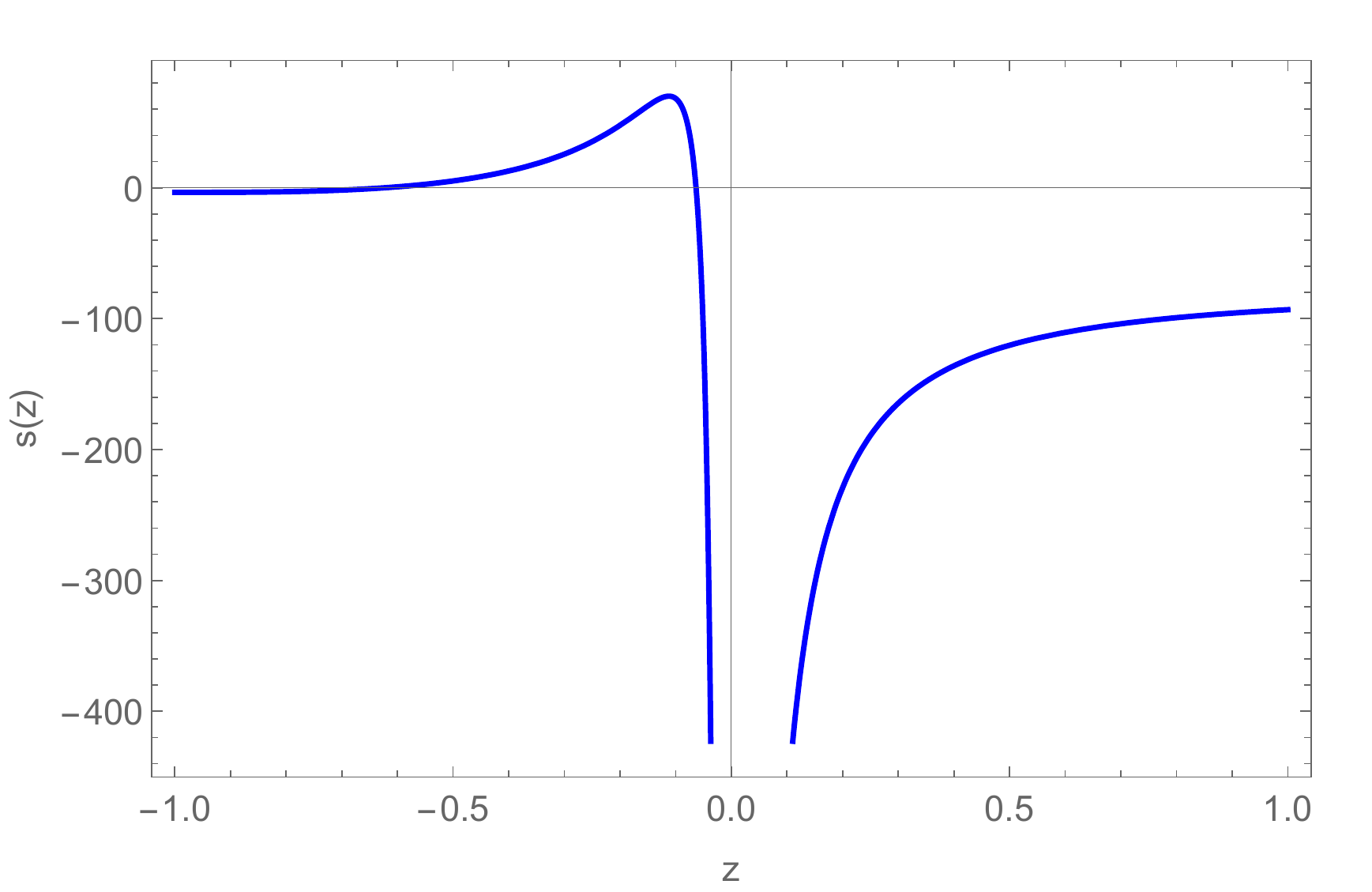}}
    \subfigure[]{\includegraphics[width=4.2cm,height=3.5cm]{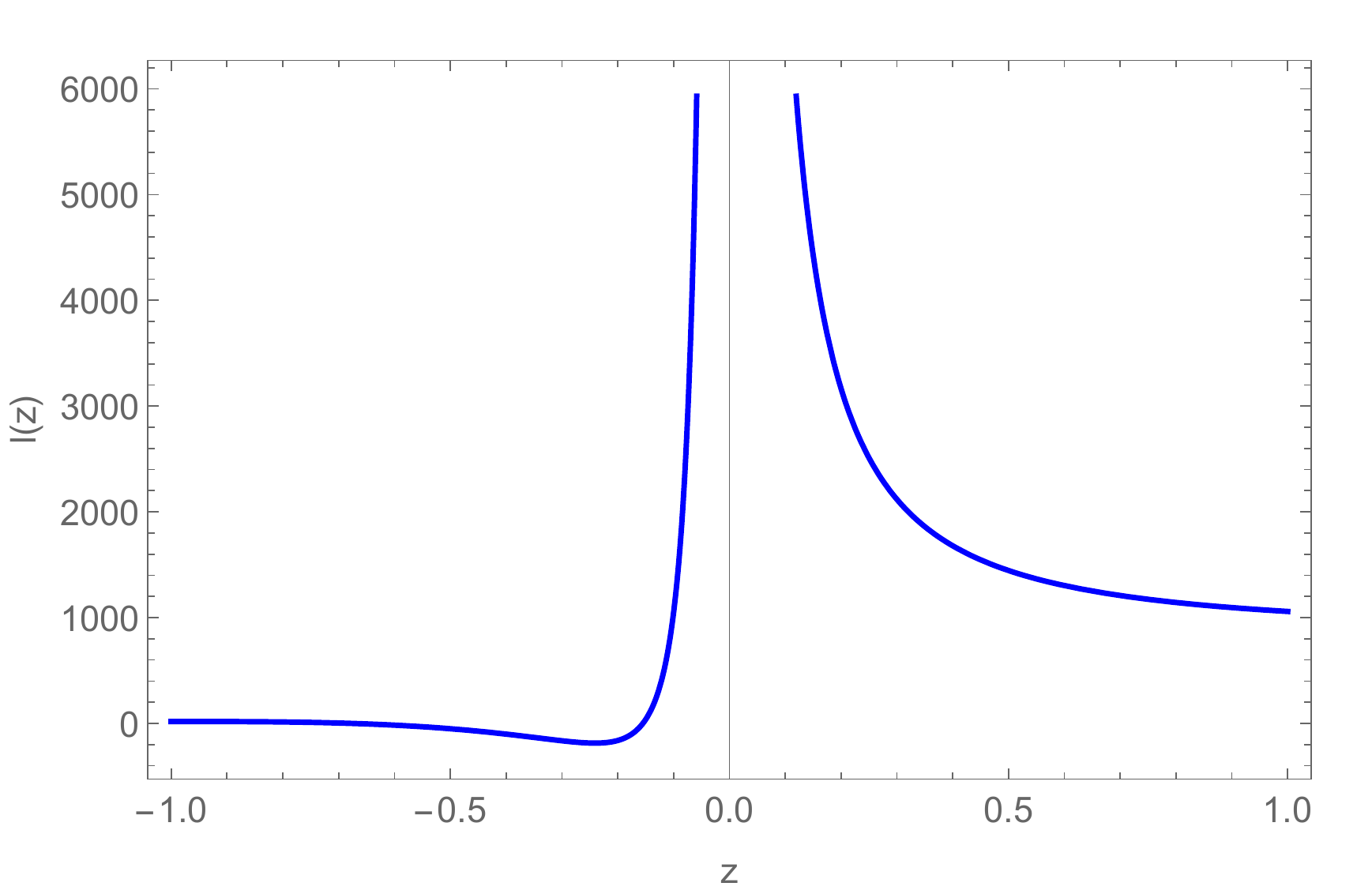}}
    \caption{Plot for the variation of (a) the deceleration parameter, (b) jerk parameter, (c) snap parameter  and (d) the lerk parameter as functions of  redshift for Model II with $\rho_{c}=0.75$.~~[Eq.\eqref{eq.29}]}
\end{figure}
\section{Stability Analysis}
In this section, we wish to study the stability of the modified $f(Q, T)$ gravity as presented. The Universe is conceived to be  filled with a perfect fluid for which we may introduce the adiabatic speed of sound $C_{s}^{2}=\frac{dp}{d\rho}$. Because the sound velocity $C_{s}^{2}$ should remain positive in a thermodynamically or mechanically stable system, the stability occurs for a positive value of $C_{s}^{2}$. Also, in order to ensure a mechanical stability, $C_{s}^{2}$ should not be larger than 1. Therefore, the region bounded by $0\leq C_{s}^{2}\leq 1$ provides stable solutions. However, when the NEC is violated, there is possibility of formation of ghost fields that suggests dangerous instabilities at the classical and/or quantum levels \cite{Dubovsky06}. Also the occurrence of superluminality may not be completely avoided. It has been shown earlier that, a regular bounce with spatial sections can only occur if the null energy condition $\rho+ p > 0$ is violated in the purview of an extended gravity theory \cite{SKT2019, SKT2020, Tripathy21}. This is the primary motivation for using negative energy scalar fields with ghost condensates \cite{Arkani-Hamed04}, conformal galileon \cite{Elder14} and other techniques to perform bounces, occasionally resulting in instabilities that must be addressed \cite{Creminelli06}. Now to separate the equations with respect to redshift \eqref{eq.15}, \eqref{eq.16} and \eqref{eq.23}, \eqref{eq.24}, we use $C_{s}^{2}$ in terms of redshift parameter to get $C_{s}^{2}$ features. 

For Model I we get, 
\begin{equation} \label{eq.30}
C_{s}^{2}=\frac{2\alpha^2\left((6\tilde{\lambda}_2-16\pi)m\chi-3\tilde{\lambda}_2\right)-\frac{\alpha\chi\left[(3\tilde{\lambda}_2-8\pi)(4m+1)\chi-9\tilde{\lambda}_2\right]}{(1+z)^{2\chi}}+\frac{\chi^{2}(\left(3\tilde{\lambda}_2-8\pi\right)m\chi-3\tilde{\lambda}_2)}{(1+z)^{4\chi}}}{2\alpha^2(2\lambda_2\chi+3\tilde{\lambda}_2)-\frac{\alpha\chi((4m+1)\lambda_2\chi+9\tilde{\lambda}_2)}{(1+z)^{2\chi}}+\frac{\chi^2(\lambda_2\chi m+3\tilde{\lambda}_2)}{(1+z)^{4\chi}}},
\end{equation}

and for the Model II, we obtain
\begin{equation}\label{eq.31}
C_{s}^{2}=\frac{\lambda_2\left[3m(2z^{3}+6z^{2}+6z+1)^2-\frac{(z^{3}+3z^{2}+3z)}{(4z^{3}+12z^{2}+12z+5)^{-1}}-3\right]+16\pi\left[m(2z^{3}+6z^{2}+6z+1)^2-\frac{(z+1)^3}{(2z^{3}+6z^{2}+6z)^{-1}}-1\right]}{\lambda_2\left[m(2z^{3}+6z^{2}+6z+1)^2+\frac{(z^{3}+3z^{2}+3z)}{(4z^{3}+12z^{2}+12z+1)^{-1}}-1\right]+16\pi\frac{[2(z^{3}+3z^{2}+3z)+1]}{(z^{3}+3z^{2}+3z)^{-1}}}.
\end{equation}

In FIG. 14 and FIG. 15, the stability of the extended symmetric teleparallel gravity models are shown. In case of Model I and II the stability condition is not satisfied as the value of $C_{s}^{2}$ remains negative. In view of this, the models may post some kind of instabilities. In order to avoid instabilities, we need $C_{s}^{2}>0$. As a result, the null energy condition can be violated while still maintaining stability.   

\begin{figure}[tbph]
\centering
\minipage{0.50\textwidth}
\includegraphics[width=\textwidth]{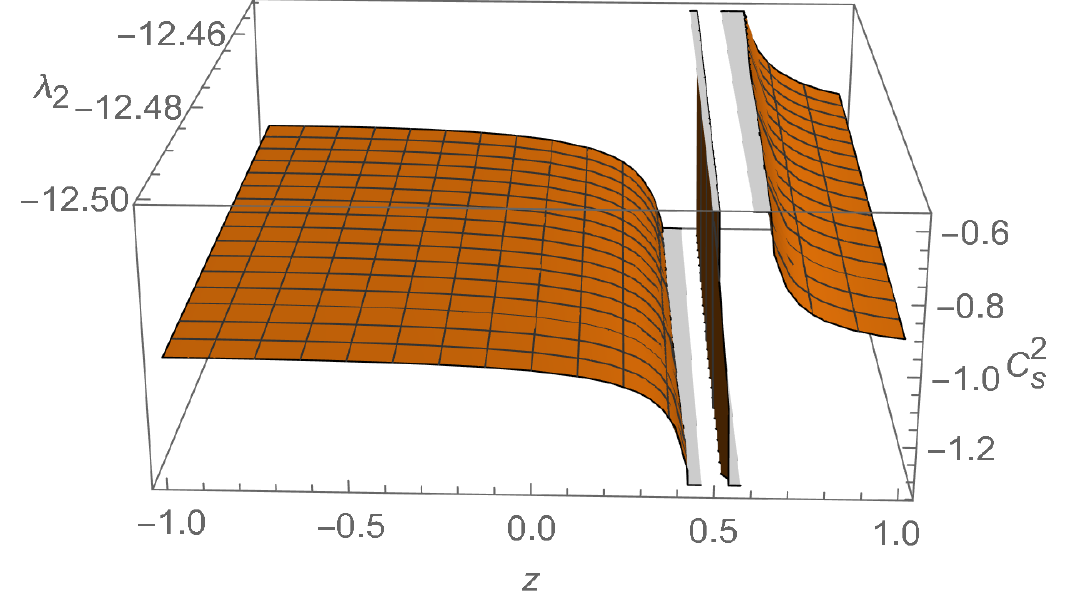}
\endminipage
\caption{Stability analysis for Model I  with the parameter space $\alpha=0.43$, $\chi=1.001$, $\lambda_1=-0.5$, $\lambda_2=-12.5$ and $m=1.01$.~~[Eq. \eqref{eq.30}]}
\end{figure}

\begin{figure}[tbph]
\minipage{0.50\textwidth}
\centering
\includegraphics[width=\textwidth]{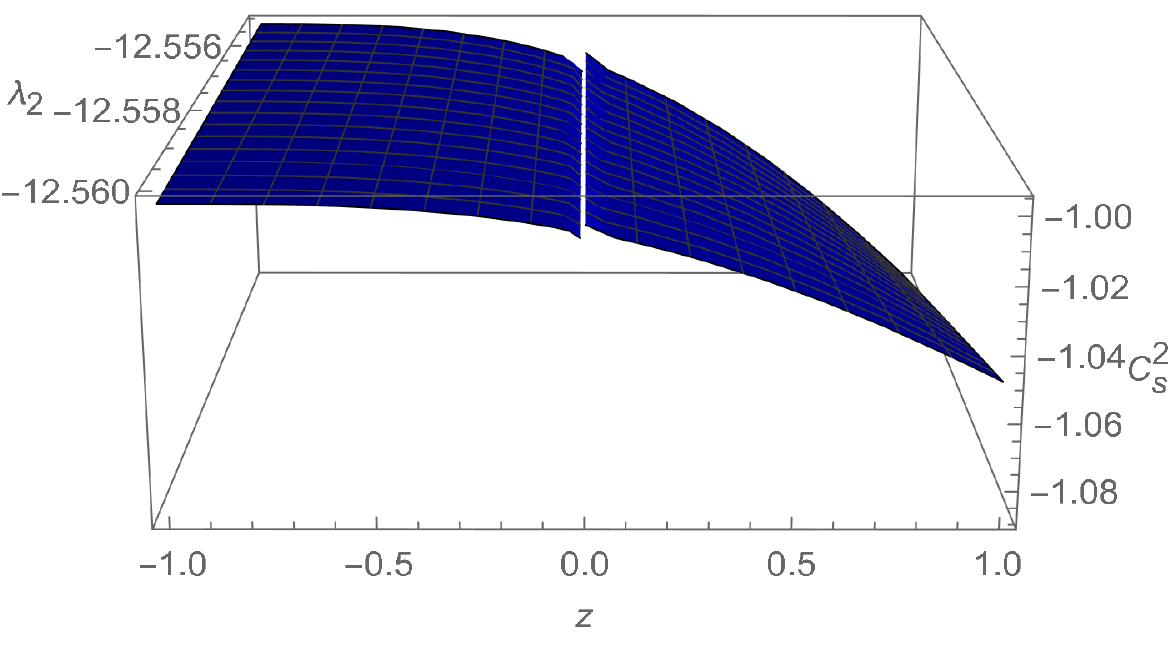} 
\endminipage
\caption{Stability analysis  for Model II with the parameter space $\rho_{c}=0.75$, $\lambda_1=-0.5$, $\lambda_2=-12.56$ and $m=1.01$.~~[Eq.\eqref{eq.31}]}
\end{figure}
\section{Conclusion}
In the present work, we have discussed the role of an extended gravity theory in providing viable models of the Universe concerning the late time acceleration issue. In order to handle the dark energy issue, many new gravity theories are proposed in literature and it is interesting to study the viability of those newly proposed gravity theories. In this context, we chose a recently proposed gravity theory dubbed as $f(Q, T)$ gravity \cite{Xu19}. In fact, the $f(Q,T)$ gravity is an extension of the symmetric parallel gravity theory, $f(Q)$ gravity. And as has been claimed in Ref.\cite{Xu19}, the $f(Q,T)$ gravity is more or less behaving the same way as $f(R,T)$ gravity. The late time acceleration issue is tackled through the geometric modification of the Einstein-Hilbert action which provides a geometrical form for the exotic dark energy. We consider a generic functional form $f(Q,T)=\lambda_1Q^m+\lambda_2T$, where $Q$ is the non-metricity function. In usual practice in literature, the EoS parameter is considered as either constant or in some parametrized form to handle the highly non linear equations of motion and then the scale factor is obtained. In that process, the dynamical behaviour of the model is constrained through the model parameters chosen which limits the evolutionary behaviour of the EoS parameter and other dynamical properties such as the energy density and pressure. In the present work, we adopted a different approach. We consider certain assumed ansatz for the scale factor and derived the expressions for the dynamical properties such as the pressure, energy density and the EoS parameter. In this process, the role of the modified gravity in the dynamical aspect of the model can be clearly assessed. We have considered two different ansatz for the scale factors that describe non-singular matter bounce scenario at some early epoch. The role of the extended symmetric teleparellel gravity on the dynamics of the  model are analysed. Through a detailed analysis, it is observed that, matter bounce scenario are possible within the formalism of $f(Q,T)$ gravity theory besides providing a suitable behaviour for the EoS parameter. The evolutionary aspect of the dynamical parameters such as the pressure, energy density and the EoS parameter show a kind of ditch/hump behaviour near the bounce. This behaviour depends on the choice of the $f(Q,T)$ parameters.

In order to justify the models, we have  validated the models through the calculation of the cosmographic parameters and the energy conditions. As it should be for any dark energy models, the strong energy condition is observed to be violated through the cosmic evolution both in the positive and negative time frame. Also, for bouncing scenario to be materialised, the null energy condition be violated. Although, we obtain marginal violation of the NEC for our models, we fix up the responsibility  to the choice of the model parameters. As such, in bouncing models, the violation of NEC leads to thermodynamical and mechanical instability in the models. In view of the results, we hope that, the new version of the symmetric telleparellel gravity may provide suitable geometrical alternatives to dark energy models. However, further investigations in this direction is required to understand the stability and the  issue of non-conservation of the energy-momentum in this geometric theory.

\section*{Acknowledgement} ASA acknowledges the financial support provided by University Grants Commission (UGC) through Senior Research Fellowship (File No. 16-9 (June 2017)/2018 (NET/CSIR)), to carry out the research work. LP acknowledges Department of Science and Technology (DST), Govt. of India, New Delhi for awarding INSPIRE fellowship (File No. DST/INSPIRE Fellowship/2019/IF190600) to carry out the research work. ASA, LP, BM acknowledge DST, New Delhi, India for providing facilities through DST-FIST lab, Department of Mathematics, where a part of this work was done. SKT and BM acknowledge the support of IUCAA, Pune (India) through the visiting associateship program.

\end{document}